\documentclass[twocolumn,times]{aastex7_sihao}

\usepackage{CJK}
\usepackage{booktabs} % For commands used in the tables
\usepackage{xcolor,soul}
\usepackage{amsmath,amssymb,bm}

\received{August 6, 2024}
\revised{May 7, 2025}
\accepted{May 23, 2025}

\submitjournal{AJ}

\shorttitle{A Candidate Giant Planet orbiting the Ember of a B Star}
\shortauthors{Cheng et al.}

\begin{document}
\begin{CJK*}{UTF8}{gkai}

\title{A Candidate Giant Planet Companion to the Massive, Young White
Dwarf GALEX J071816.4+373139 Informs the Occurrence of Giant Planets
Orbiting B Stars}

\correspondingauthor{Sihao Cheng}

\author[0000-0002-9156-7461]{Sihao Cheng (程思浩)}
%\author[0000-0002-9156-7461]{Sihao Cheng}
\affiliation{Institute for Advanced Study, 1 Einstein Dr, Princeton,
NJ 08540, USA}
\email[show]{scheng@ias.edu}

\author[0000-0001-5761-6779]{Kevin C.\ Schlaufman}
\affiliation{William H.\ Miller III Department of Physics \& Astronomy,
Johns Hopkins University, 3400 N Charles St, Baltimore, MD 21218, USA}
\email{kschlaufman@jhu.edu}

\author[0000-0002-4770-5388]{Ilaria Caiazzo}
\affiliation{Institute of Science and Technology Austria, Am Campus 1, 
3400 Klosterneuburg, Austria}
\affiliation{Division of Physics, Mathematics and Astronomy, California
Institute of Technology, 1200 E California Blvd, Pasadena CA 91125, USA}
\email{Ilaria.Caiazzo@ist.ac.at}

\begin{abstract}

\noindent
It has been suggested that giant planet occurrence peaks for stars
with $M_{\ast}~\approx~3~M_{\odot}$ at a value a factor of 4 higher
than observed for solar-mass stars.  This population of giant planets
predicted to frequently orbit main-sequence B stars at $a~\approx$ 10au
is difficult to characterize during the few hundred million years while
fusion persists in their host stars.  By the time those stars become
massive, young white dwarfs, any giant planets present would still be
luminous as a consequence of their recent formation.  From an initial
sample of 2195 Gaia-identified massive, young white dwarfs, we use
homogeneous Spitzer Infrared Array Camera (IRAC) photometry to search
for evidence of unresolved giant planets.  For 30 systems, these IRAC
data provide sensitivity to objects with $M~\lesssim~10~M_{\text{Jup}}$,
and we identify one candidate with $M~\approx~4~M_{\text{Jup}}$ orbiting
the white dwarf GALEX J071816.4+373139.  Correcting for the possibility
that some of the white dwarfs in our sample result from mergers, we find
a giant planet occurrence $\eta_{\text{GP}}~=~0.11_{-0.07}^{+0.13}$ for
stars with initial masses $M_{\ast}~\gtrsim~3~M_{\odot}$.  Our occurrence
inference is consistent with both the Doppler-inferred occurrence of
giant planets orbiting $M_{\ast}~\approx~2~M_{\odot}$ giant stars and the
theoretically predicted factor of 4 enhancement in the occurrence of
giant planets orbiting $M_{\ast}~\approx~3~M_{\odot}$ stars relative to
solar-mass stars.  Future James Webb Space Telescope NIRCam
observations of our sample would provide sensitivity to Saturn-mass
planets and thereby a definitive estimate of the occurrence of giant
planets orbiting stars with $M_{\ast}~\gtrsim~3~M_{\odot}$.
\end{abstract}


\section{Introduction}\label{sec:intro}

The relationship $\eta_{\text{GP}}(M_{\ast},Z)$ between giant
planet occurrence $\eta_{\text{GP}}$ and the masses $M_{\ast}$
and metallicities $Z$ of their host stars has provided some of the
strongest evidence in support of the core-accretion model of giant
planet formation \citep[e.g.,][]{Pollack_1996, Hubickyj_2005}.
For dwarf stars in the interval $0.3~M_{\odot} \lesssim M_{\ast}
\lesssim 1.0~M_{\odot}$, $\eta_{\text{GP}}(M_{\ast})$ is correlated with
$M_{\ast}$ \citep[e.g.,][]{Butler_2006,Cumming_2008}.  These observations
confirmed theoretical predictions put forward by \citet{Laughlin_2004} and
\citet{Ida_2005}.  Likewise, $\eta_{\text{GP}}(M_{\ast})$ increases with
$M_{\ast}$ in the interval $1.0~M_{\odot} \lesssim M_{\ast} \lesssim
2.0~M_{\odot}$ with the caveat that $\eta_{\text{GP}}(M_{\ast})$
inferences are based on dwarf stars at the low end of that interval
and evolved stars with much larger stellar pulsations at the high
end of that interval \citep[e.g.,][]{Johnson_2010, Ghezzi_2018,
Wolthoff_2022}.  $\eta_{\text{GP}}(M_{\ast},Z)$ increases with $Z$
over the mass interval $0.3~M_{\odot} \lesssim M_{\ast} \lesssim
2.0~M_{\odot}$, with some suggestion that the strength of the dependence
of $\eta_{\text{GP}}(M_{\ast},Z)$ on $Z$ is diminished at the high-mass end of that interval \citep[e.g.,][]{Santos_2004, Fischer_2005,
Hekker_2007, Pasquini_2007, Takeda_2008, Maldonado_2013, Reffert_2015,
Jones_2016, Wittenmyer_2017, Wolthoff_2022}.

Taken together, the observations that $\eta_{\text{GP}}(M_{\ast},Z)$
increases with both $M_{\ast}$ and $Z$ confirms the key prediction of
the core-accretion model of giant planet formation that cores of giant
planets can assemble more quickly in disks with larger complements of
solid material.  Gaseous disks orbiting young stars are observed to
dissipate more quickly for massive stars, though, so while the timescale
for core formation decreases for massive stars, so does the time
available for giant planet formation \citep[e.g.,][]{Carpenter_2006,
Kennedy_2009, Roccatagliata_2011, Yasui_2014, Ribas_2015}.  The net
result of these two effects is that at constant metallicity the
function $\eta_{\text{GP}}(M_{\ast})$ should peak somewhere in
the interval $M_{\ast} \gtrsim 1~M_{\odot}$, with the exact value
sensitively dependent on the details of the planet formation process
\citep[e.g.,][]{Ida_2005,Kennedy_2008, Kretke_2009}.  The measurement of
$\eta_{\text{GP}}(M_{\ast})$ for the most massive stars with $M_{\ast}
\gtrsim 3~M_{\odot}$ would therefore provide critical empirical
constraints on the giant planet formation process.

The observational characterization of the giant planet population orbiting
stars with $M_{\ast} \gtrsim 3~M_{\odot}$ is extraordinarily difficult
with the Doppler and transit techniques responsible for the discovery
and characterization of most known exoplanets.  Massive stars have
large-amplitude pulsations, fast rotation, and few absorption lines,
all of which make it very hard to discover and characterize planets
with the Doppler technique.  The pulsations and large radii of massive
stars complicate searches for planets using the transit technique, too,
and the Doppler confirmation of candidate transiting planets faces the
same problems.  Because the water-ice lines in the massive protoplanetary
disks initially present around $M_{\ast} \gtrsim 3~M_{\odot}$ stars
will lie beyond 10 au, the relatively wide-separation planets predicted
to form would be invisible to the Doppler or transit techniques, even
without the difficulties described above.

The expected peak of the giant planet population orbiting stars
with $M_{\ast} \gtrsim 3~M_{\odot}$ is also inaccessible with direct
imaging techniques.  Since solar metallicity stars with $M_{\ast} \gtrsim
3~M_{\odot}$ leave the main-sequence in less than about 400 Myr, main
sequence stars with $M_{\ast} \gtrsim 3~M_{\odot}$ must be young and
therefore attractive targets for the direct imaging technique.  Massive
stars are uncommon, though, as only 18 such stars are present in the Gaia
Early Data Release 3 (EDR3) 100 pc sample \citep{GaiaCollaboration_2021b}.
Those 18 massive stars have a median distance in excess of 90 pc and
$H \approx 4.5$.  Next-generation extreme adaptive optics systems
on ground-based eight-meter class telescopes like the Gemini Planet
Imager 2.0 \citep[GPI 2.0;][]{Chilcote_2022} operating in ideal
conditions will be able to achieve $H$-band contrasts of $10^{7}$
at separations of 0.2\arcsec.  According to the Sonora grid of
evolutionary models presented in \citet{Marley_2021}, a 400 Myr old
$M_{\text{p}} \approx 10~M_{\text{Jup}}$ planet will have $H \approx
23$.  At a distance of 90 pc orbiting a $M_{\ast} \approx 3~M_{\odot}$
star at semimajor axis $a \approx 10$ au close to the peak of the
occurrence of such objects orbiting $M_{\ast} \approx 2~M_{\odot}$
stars \citep[e.g.,][]{Nielsen_2019, Wolthoff_2022}, a $H$-band direct
imaging detection of that planet would require a contrast of $10^{10}$
at less than 0.2\arcsec.

As a consequence, even next-generation ground-based instruments
will be insensitive to planet-mass objects orbiting massive
stars at the separations where theoretical predictions plus
long-term Doppler monitoring and direct imaging observations
of $M_{\ast} \approx 2~M_{\odot}$ stars suggest they should be
present \citep[e.g.,][]{Reffert_2015, Nielsen_2019, Wolthoff_2022}.
Gaia astrometry will also be insensitive to giant planets orbiting stars
with $M_{\ast} \gtrsim 3~M_{\odot}$, as the few nearby stars in this mass
range saturate Gaia's detectors and make the most precise astrometric
measurements impossible.  For all of these reasons, only a few objects
in the NASA Exoplanet Archive \citep{Akeson_2013} orbit stars with
$M_{\ast} \gtrsim 3~M_{\odot}$ \citep[e.g.,][]{Janson_2019, Bang_2020,
Squicciarini_2022}.  Though not listed in the NASA Exoplanet Archive,
the B-star Exoplanet Abundance Study (BEAST) has also found objects those
authors refer to as planets \citep{Janson_2021, Janson_2021nature}.  None
of these few objects have $M \lesssim 10~M_{\text{Jup}}$ and $a \approx
10$ au as expected for objects frequently formed via core-accretion
for stars in this mass range near their parent protoplanetary disks'
water-ice lines \citep{Kennedy_2008}.

\begin{figure*}
    \centering
    \includegraphics[width=\columnwidth]{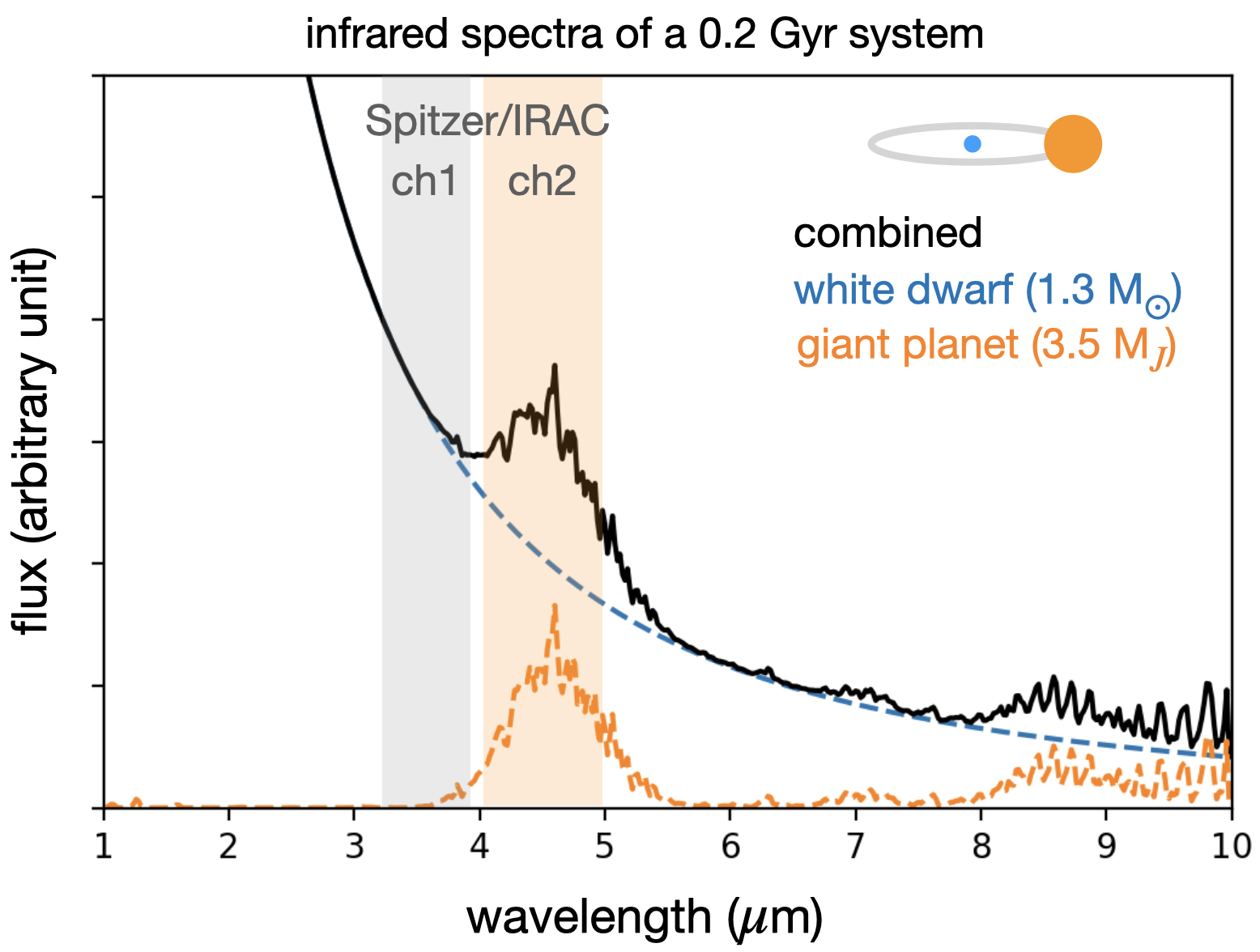}
    \includegraphics[width=0.93\columnwidth]{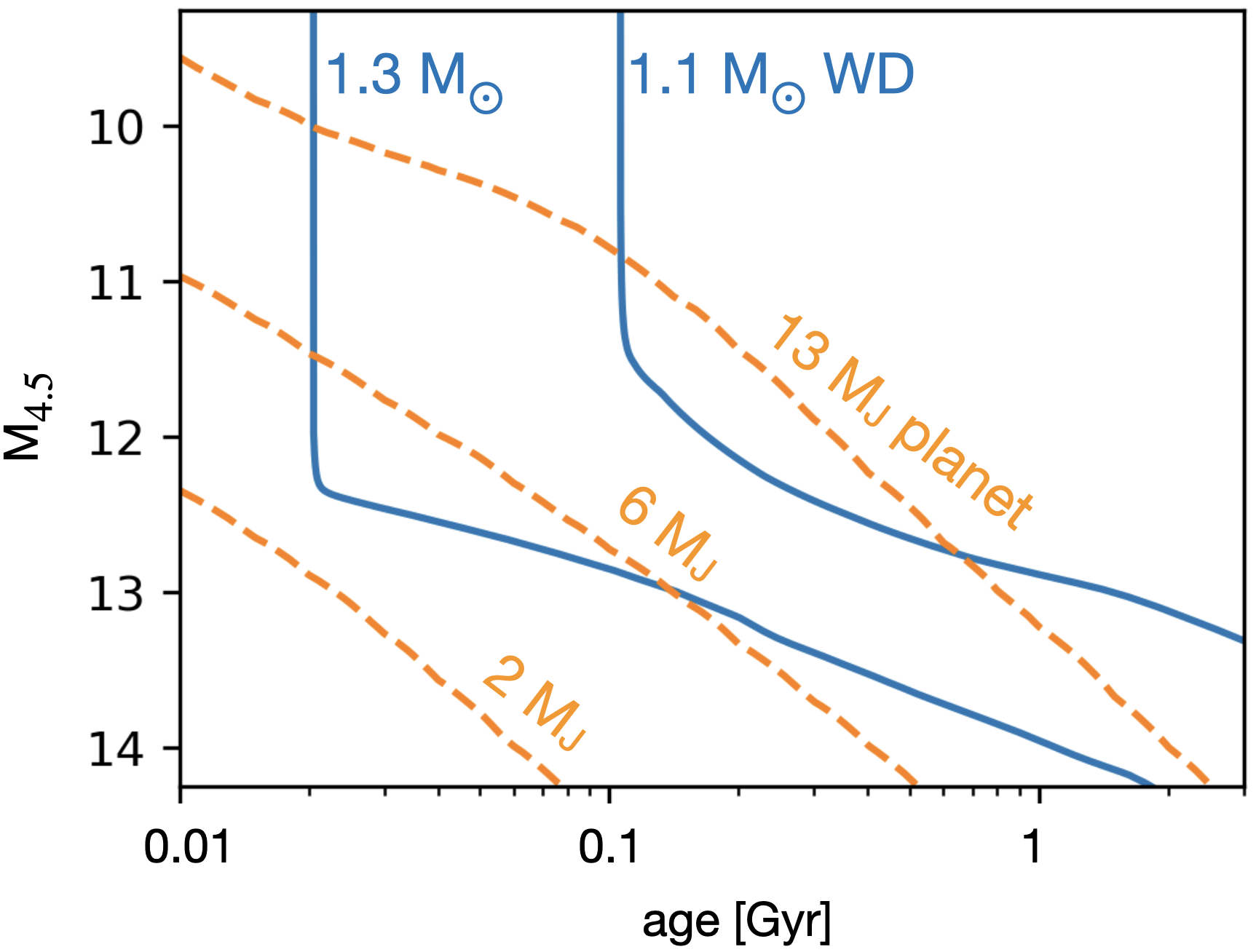}
    \caption{Left: composite spectrum of a typical young white dwarf
    plus young giant plant system.  Right: absolute magnitude $M_{4.5}$
    in the Spitzer Infrared Array Camera (IRAC) 4.5-micron Channel 2 band
    as a function of time for both massive white dwarfs and giant planets.
    Planet-mass objects with clear atmospheres are therefore observable
    for at least a few tens of Myr after their host stars become white
    dwarfs.}
    \label{fig:IR+cooling}
\end{figure*}

While this population of giant planets predicted to frequently orbit
stars with initial mass $M_{\ast,i} \gtrsim 3~M_{\odot}$ (i.e., main
sequence B stars) at $a \approx 10$ au is difficult to characterize
while fusion persists in their host stars, the population can be studied
with the direct imaging technique once their hosts become massive, young
white dwarfs \citep{Burleigh_2008, Hogan_2009, Hogan_2010, Hogan_2011,
Xu_2015, Brandner_2021}.  Despite the fact that young white dwarfs
are hot, their radii would be an order of magnitude smaller than the
radii of any equally young giant planet-mass objects that might be
present.  In that situation, depicted in Figure \ref{fig:IR+cooling},
as a consequence of an opacity minimum in young giant planet-mass
objects' cloudless atmospheres, such objects would be observable as
significant infrared excesses at 4.5 $\mu$m with only the expected
fluxes from the host white dwarf's photosphere at 3.6 or 5.8 $\mu$m
\citep[e.g.,][]{Burrows_1997, Burleigh_2002, Gould_2008, Marley_2021}.
Excess flux above the expected contribution from the white dwarf's
photosphere at 4.5 $\mu$m combined with a lack of excess flux at 3.6
and 5.8 $\mu$m is difficult to reproduce with unresolved emission from a
disk of dust or contamination by a background galaxy.  This signature is
consequently an excellent way to identify candidate giant planets orbiting
massive, young white dwarfs that were B stars on the main sequence.

Although previous searches for Spitzer Infrared Array Camera (IRAC) unresolved mid-infrared excess
indicative of the presence of giant planets orbiting solar-neighborhood
white dwarfs produced nondetections \citep[e.g.,][]{Mullally_2007,
Farihi_2008, Kilic_2009}, the unprecedented resolution and photometric
precision of the James Webb Space Telescope (JWST) Mid-infrared Instrument
(MIRI) have recently enabled the discovery of resolved giant planet
candidates orbiting both nearby white dwarfs and white dwarfs with
metal-enriched photospheres \citep{ Limbach_2024, Mullally_2024}.

In this article, we describe a search for unresolved infrared excess
in a sample of 51 massive, young white dwarfs with homogeneously
reduced archival Spitzer/IRAC [3.6] and [4.5] photometry.  These data
provide sensitivity to giant planet-mass objects with $M \lesssim
10~M_{\text{Jup}}$ for 30 white dwarfs.  In this sample, we identify one
candidate giant planet in the GALEX J071816.4+373139 white dwarf system
and infer the occurrence of giant planets orbiting main-sequence B stars
with $M_{\ast,i} \gtrsim 3~M_{\odot}$.  We outline the construction of
our sample and its analysis in Section \ref{sec:data}.  We then focus on
the properties of the giant planet candidate in the GALEX J071816.4+373139
system in Section \ref{sec:cand} and the inference of $\eta_{\text{GP}}$
in Section \ref{sec:pop}.  We conclude by summarizing our results in
Section \ref{sec:conc}.

\section{Data and Analysis}\label{sec:data}

\subsection{Selection of massive, young white dwarfs}

We use the publicly available \texttt{WD\_models} code
\citep{WD_models}\footnote{\url{https://github.com/SihaoCheng/WD\_models}}
described in detail in \citet{Cheng_2019} to construct a solar
neighborhood sample of massive, young white dwarfs suitable for the
detection of unresolved giant planets using Spitzer/IRAC photometry.
We first infer effective temperatures $T_{\text{eff}}$ by fitting an
interpolated grid of synthetic photometry predicted to be produced by
both DA (hydrogen-rich) and DB (helium-rich) white dwarf model atmospheres \citep{Holberg_2006,
Kowalski_2006, Tremblay_2011, Bergeron_2011, Blouin_2018} to
observed Gaia Data Release 3 (DR3) $G_{\text{BP}}-G_{\text{RP}}$
colors and $M_{G}$ absolute magnitudes \citep{GaiaCollaboration_2016,
GaiaCollaboration_2021a, Fabricius_2021, Riello_2021, Rowell_2021,
Torra_2021}.  With $T_{\text{eff}}$ in hand, we then infer white dwarf
radii using observed fluxes, Gaia DR3 parallax-based distance inferences
\citep{Lindegren_2021a, Lindegren_2021b}, and the Stefan--Boltzmann
law.  Assuming that the white dwarfs in our sample were the result of
the evolution of single stars, we next use the white dwarf cooling
models presented in \citet{Bedard_2020} to infer white dwarf masses
$M_{\text{WD}}$ and cooling ages $\tau_{\text{cool}}$.  We finally
calculate white dwarf total ages $\tau = \tau_{\ast} + \tau_{\text{cool}}$
based on the \citet{Cummings_2018} zero-age main sequence to white dwarf
initial--final mass relation.  While the assumption of single stellar
evolution and modeling uncertainties are inherent in this approach,
the white dwarf cooling models we use are well calibrated and in accord
with other white dwarf cooling models \citep{Salaris_2013}.  Since age
inference uncertainties for massive white dwarfs are dominated by cooling
age uncertainties, the system ages we subsequently use in our sample
selection and search for unresolved giant planets should be robust.

\begin{figure}
    \centering
    \includegraphics[width=\columnwidth]{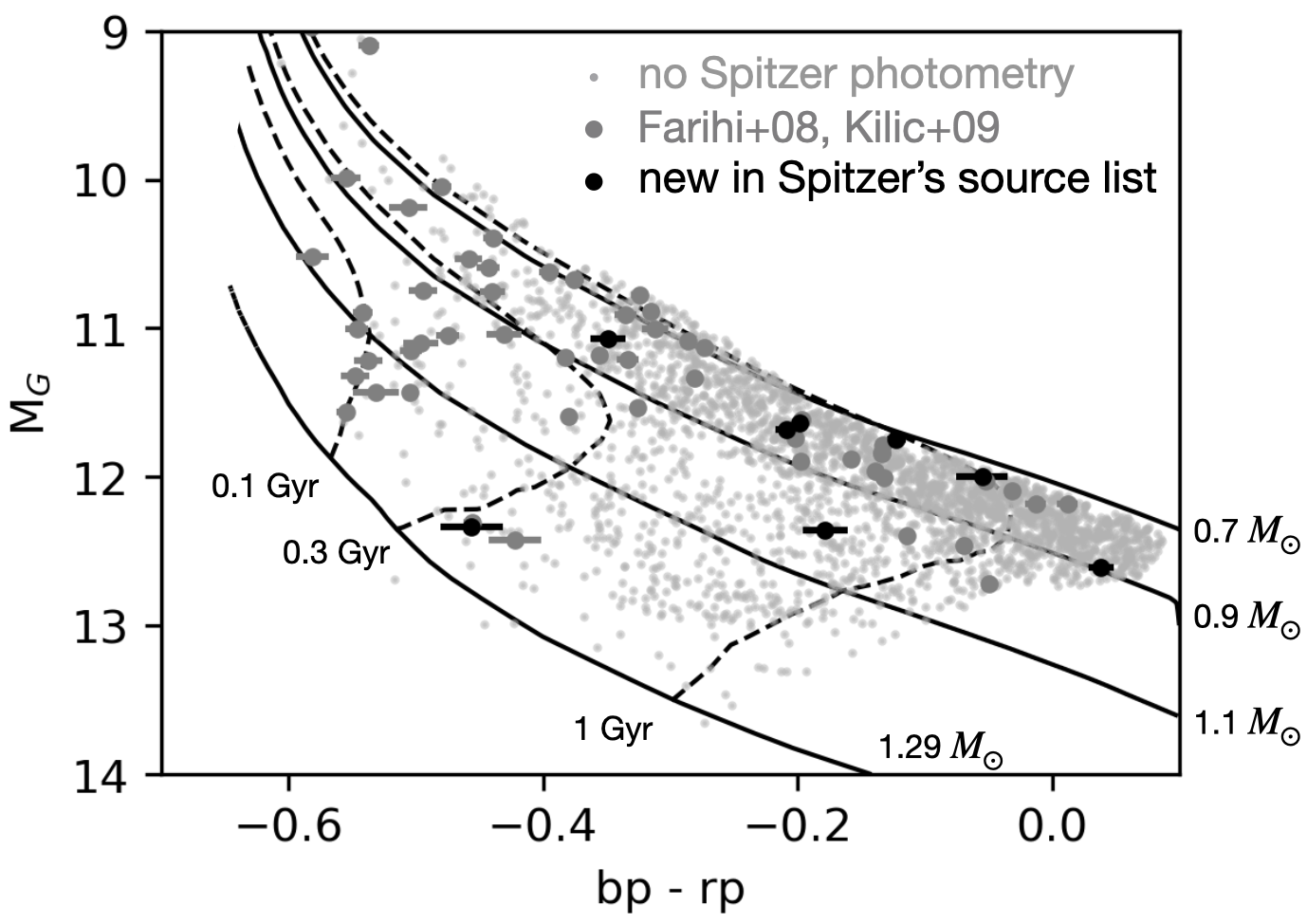}
    \caption{HR diagram of massive, young white dwarfs in our sample.
    We plot our sample of 2195 white dwarfs as the smaller gray points.
    We plot as larger gray points white dwarfs that were searched for
    the infrared excess characteristic of unresolved giant planets by
    \citet{Farihi_2008} or \citet{Kilic_2009}.  We plot as black points
    white dwarfs in our sample with SEIP Source List photometry searched
    for the infrared excess characteristic of unresolved giant planets
    for the first time in this article.  We plot as yellow curves cooling
    tracks for DA white dwarfs with $M_{\text{WD}} = (0.7, 0.9, 1.1,
    1.29)~M_{\odot}$.}
    \label{fig:HR}
\end{figure}

We apply the procedure described above to Gaia DR3 objects identified
as white dwarfs by \citet{Gentile_Fusillo_2019}.  In particular, we
select objects with $\tau < 1.2$ Gyr, $M_{\text{WD}} > 0.7~M_{\odot}$,
and $d = 1/\pi < 200$ pc.  This procedure leads to the sample of 2195
massive, young white dwarfs in the solar neighborhood illustrated in
Figure \ref{fig:HR}.

\subsection{Spitzer/IRAC analysis of our white dwarf sample}

To search for the infrared excess characteristic of unresolved
giant planets orbiting our white dwarf sample, we crossmatch our
sample with the Spitzer Enhanced Imaging Products (SEIP) Source
List\footnote{\url{https://doi.org/10.26131/IRSA433}} \citep{https://doi.org/10.26131/irsa433} available from the
NASA/IPAC Infrared Science Archive (IRSA).  The SEIP Source List includes
homogeneous photometric measurements for all point-like sources in areas
of the sky observed from 2003 to 2009 during Spitzer's cryogenic phase.
We account for the proper motions of the white dwarfs in our catalog
by first converting the epoch 2016.5 Gaia DR3 right ascensions and
declinations to the mean epoch of the Spitzer cryogenic phase 2006.0.
We then match the white dwarfs in our catalog with the nearest SEIP
Source List object within 1.5\arcsec.

This procedure results in the 65 matches with IRAC photometry illustrated
in Figure \ref{fig:HR}, 51 with both [3.6] and [4.5] plus 14 with only
[4.5].  Of these 65 matches, 57 were observed by white dwarf-focused
programs and have been previously studied \citep{Mullally_2007,
Farihi_2008, Kilic_2009, Barber_2016}.  The other eight were
serendipitously observed by Spitzer programs that were not focused on
white dwarfs.  While a larger absolute number of white dwarfs were
studied by \citet{Wilson_2019} using data from Spitzer's warm phase
(Programs 80149 and 12103), only one of our massive, young white
dwarfs WD 1134+300 was included in that study.  We note that even for
white dwarfs studied previously, SEIP Source List photometry provides
better accuracy and higher precision than previous studies due to its
production by the most up-to-date pipeline using the best zero-point
and photometric corrections calibrated with all Spitzer cryogenic data
available.\footnote{\url{https://irsa.ipac.caltech.edu/data/SPITZER/Enhanced/SEIP/docs/seip\_explanatory\_supplement\_v3.pdf}}
In particular, as we will describe in detail in Section \ref{sec:color
excess}, these SEIP Source List data warrant a reanalysis because residual
$\chi^2$ values suggest that published analyses of these same Spitzer
data likely overestimated photometric uncertainties.

Unlike the data sets produced by the Wide-field Infrared Explorer
\citep[WISE;][]{Wright_2010}, the Spitzer SEIP Source List is not an
all-sky catalog.  Nevertheless, Spitzer's superior sensitivity and
angular resolution relative to WISE makes the SEIP Source List a better
data set to use for a search for the infrared excess characteristic of
unresolved giant planets.  Indeed, we also explored the possibility
of using CatWISE2020 W1 and W2 photometry \citep{Mainzer_2011,
Eisenhardt_2020, Marocco_2021} as an all-sky supplement to the more
precise Spitzer SEIP Source List photometry.  Like other groups
\citep[e.g.,][]{Dennihy_2017, Xu_2020, Lai_2021}, we found that even
after removing from consideration objects with UK Infra-Red Telescope
(UKIRT) Hemisphere Survey \citep[UHS;][]{Dye_2018} or Visible and
Infrared Survey Telescope for Astronomy (VISTA) Hemisphere Survey
\citep[VHS;][]{McMahon_2013} $J$-band excesses, a large number of white
dwarfs with apparent CatWISE2020 infrared excess could still be attributed
to infrared-bright galaxies with very faint near-infrared counterparts.
This increased incidence of confusion relative to the Spitzer SEIP Source
List is a consequence of WISE's coarser spatial resolution and lower
sensitivity.  We ultimately concluded that CatWISE2020 is unsuitable for a
search for the infrared excess characteristic of unresolved giant planets.

\subsection{Color and 4.5 $\mu$m flux excesses}
\label{sec:color excess}

To quantify the statistical significance of any excess flux in the
mid infrared beyond the expected contribution from the white dwarf
photosphere, we define the flux and color excesses as
\begin{eqnarray}
\text{flux excess} & = & m_\text{model}(M_G, G_{\text{BP}} - G_{\text{RP}}) - m_\text{obs} + \delta_{m}, \\
\text{color excess} & = & C_\text{obs} - C_\text{model}(M_G, G_{\text{BP}} - G_{\text{RP}}) + \delta_{C}.
\end{eqnarray}
Here $m$ refers an apparent magnitude and $C$ refers to the quantity
$m_{[3.6]}-m_{[4.5]}$ assuming a 3.8\arcsec aperture.  $\delta_{m}$ and
$\delta_{C}$ represent the approximately 1.5\% systematic uncertainties
of Spitzer's absolute photometric calibration.  The presence of an
unresolved giant planet would manifest as a flux excess at 4.5 $\mu$m
but not at 3.6 $\mu$m.  On the other hand, a dust disk or background
galaxy would be expected produce excesses at both 3.6 and 4.5 $\mu$m.
To make optimal use of the available photometric data and thereby
obtain the best sensitivity to unresolved giant planets, we focus on
color excess between the two bands as a relative quantity with lower
uncertainty instead of excesses in individual fluxes.

The flux and color excess uncertainties have contributions from both
the input photometry itself and white dwarf model parameter uncertainties
\begin{eqnarray}
\sigma_\text{flux excess} & = & \sqrt{\sigma_{\text{obs}}^2 + \sigma_{\text{model}}^2 + \sigma_{\text{extra}}^2}, \\
\sigma_\text{color excess} & = & \sqrt{\sigma_{\text{obs,3.6}}^2 + \sigma_{\text{obs,4.5}}^2 + \sigma_\text{color,model}^2 + \sigma_{\text{color, extra}}^2}
\end{eqnarray}

The advantage of color excess is that (1) its model error is almost
zero because parallax errors cancel out between the two bands and (2)
for hot white dwarfs, color is not sensitive to temperature.  This latter
point is a consequence of the fact that in this case Spitzer data only
cover the Rayleigh-Jeans tail of each white dwarf's spectral energy
distribution that has a color index close to zero.  In contrast, flux
excess as an absolute photometric quantity has a higher uncertainty as
a consequence of photometric uncertainty, parallax uncertainty, white
dwarf temperature estimate uncertainty, and possibly extinction and its
uncertainty.\footnote{According to the formula for error propagation
\begin{eqnarray}
\sigma_{\text{model}} & = & \sqrt{(\sigma_{M_G} \cdot \partial_{M_G} m_{\text{model}})^2 + (\sigma_{G_{\text{BP}}-G_{\text{RP}}} \cdot \partial_{G_{\text{BP}}-G_{\text{RP}}} m_{\text{model}})^2}, \\
& \approx & \sqrt{\left(\frac{2.17 \sigma_{\omega}}{\omega}\right)^2 + \sigma_{G}^2 + (\sigma_{G_{\text{BP}}-G_{\text{RP}}} \cdot \partial_{G_{\text{BP}}-G_{\text{RP}}} m_{\text{model}})^2}.
\end{eqnarray}}
We therefore use color excess for the selection of giant planet candidates
and only use flux excess for the exclusion of possible contamination.
We note that the nominal uncertainty of aperture photometry in the
SEIP Source List is underestimated by a factor of 2, so we set
$\sigma_{\text{obs}}$ to be 2 times the nominal photometric uncertainty
for the 3.8\arcsec aperture in the SEIP Source List.

We also add an extra uncertainty term to include other sources of
uncertainties, such as the intra-pixel variation of IRAC photometry.
We note that similar published analyses tend to overestimate color excess
uncertainties as indicated by the fact that their dispersions of color
excess significance $\Sigma = \text{color excess} / \sigma(\text{color
excess})$ are smaller than unity: 0.8 in \citet{Wilson_2019} and 0.6 in
\citet{Lai_2021}.  Overly conservative uncertainties are not problems
for searches for large infrared excesses due to disks or brown dwarfs
that are bright and tend to produce strong signals.  However, an overly
conservative uncertainty would significantly reduce the constraining
power of a planet search.  We find that $\sigma_{\text{color, extra}} =
3\%$ and $\delta_C = -0.7\%$ lead to a zero mean and unity dispersion
in $\Sigma$ after clipping 3$\sigma$ outliers.  Similarly, for the
flux excesses $\sigma_{\text{3.6, extra}} = 5\%$, $\delta_{3.6} =
-1.5\%$, $\sigma_{\text{4.5, extra}} = 4\%$, and $\delta_{4.5} = -0.8\%$
yield zero means and unity dispersions for $\chi = \text{flux excess}
/ \sigma(\text{flux excess})$.  These values are consistent with the
values used in the literature \citep[e.g.,][]{Farihi_2008, Wilson_2019,
Lai_2021}.

\begin{figure}
    \centering
    \includegraphics[width=\columnwidth]{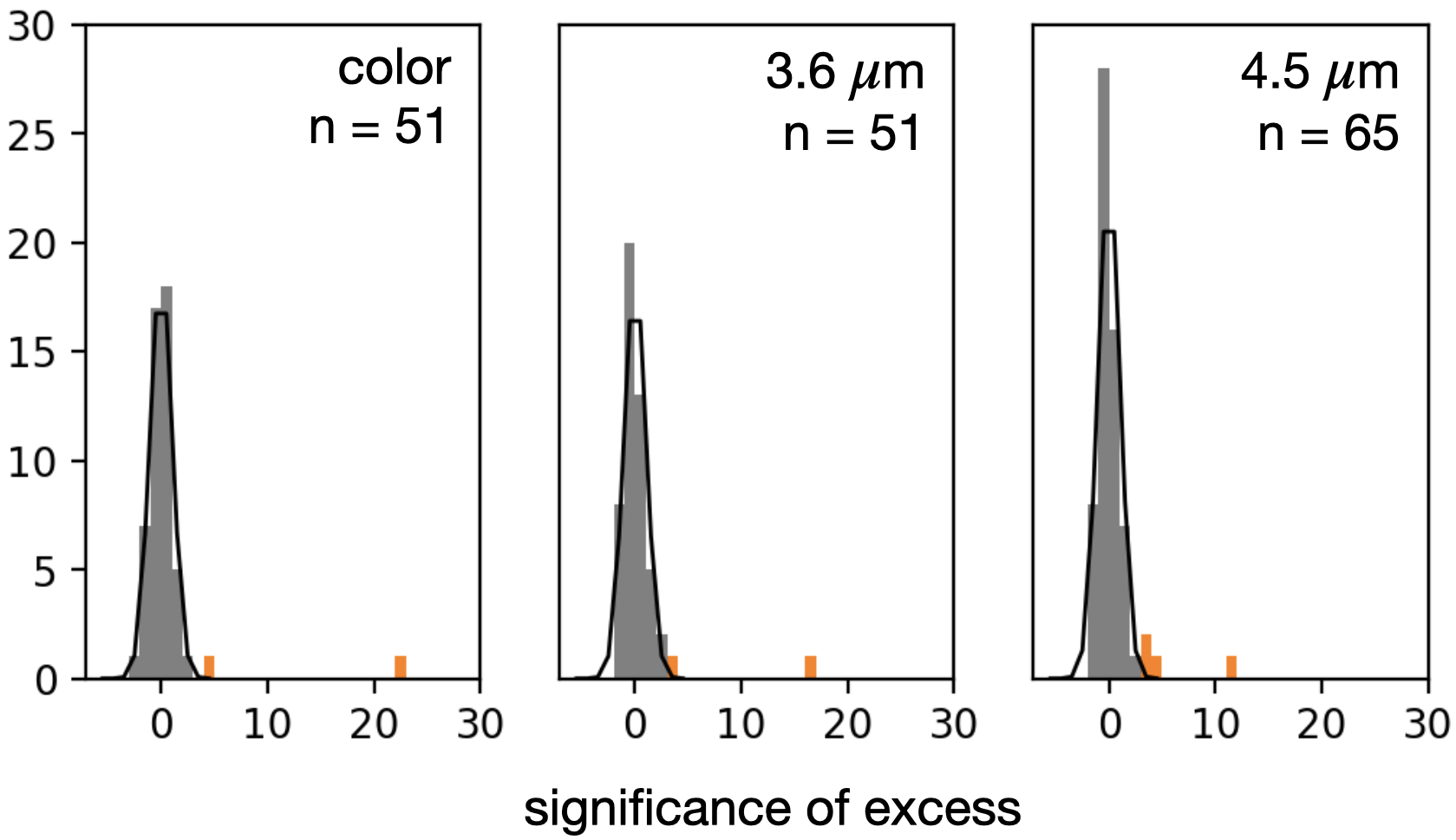}
    \caption{Histograms of infrared excess significances.  Note that
    cores of each distribution match well with a standard Gaussian profile
    plotted in black.  We indicate individual white dwarfs with excesses
    beyond $3\sigma$ in orange.} 
    \label{fig:Chi}
\end{figure}

We plot in Figure \ref{fig:Chi} histograms of $\Sigma$, $\chi_{3.6}$,
and $\chi_{4.5}$ which show core profiles that are consistent with a
standard Gaussian distributions with (mean, standard deviation) ordered
pairs (0.02, 0.98), (0.01, 1.02), and (-0.01, 1.01).  As a cross validation
of our approach to color and flux excesses, we also examined another
sample of 379 white dwarfs in the SEIP Source List that are not part of
our science sample.  After removing objects with large (30\%) 4.5 micron
flux excesses, the color excess significance $\Sigma$ has (--0.00, 1.04).
This results further validates the photometric reliability of the SEIP
Source List and our color excess uncertainty model.  It also shows
that our approach is applicable to other Spitzer observations.  On the
other hand, as argued above the flux excess is not as stable, as it has
(0.09, 1.15) and (0.11, 1.13) in [3.6] and [4.5].

\begin{table}
    \centering
    \begin{tabular}{ccccccc}
    \hline
    White dwarf & Gaia DR3 \texttt{source\_id} &  $\Sigma$ & $\chi_{3.6}$ & $\chi_{4.5}$ \\
    \hline
    0003+4336 & 384841336150015360  & 2.1  & -1.1 & 1.3 \\
    0148+1902 & 95297185335797120   & 22.4 & 52.7 & 120.6 \\
    0718+3731 & 898348313253395968  & 4.3  & -1.1 & 3.1 \\
    % 1227--0814& 3583181371265430656 & 2.1  & 2.4 & 4.8 \\
    % 1228+1040 & 3904415787947492096 & 21.8 & 82.6 & 167.2 \\
    \hline
    \end{tabular}
    \caption{White dwarfs with color excess larger than 2$\sigma$.  Most
    are consistent with a blackbody origin.  We flag as candidates only
    white dwarfs with 3$\sigma$ significant flux excess at 4.5 $\mu$m
    with both (1) no statistically significant flux excess at 3.6
    $\mu$m and (2) an absolute flux excess at 3.6 $\mu$m less than 0.5
    mag.  Among the 51 objects in our sample for which these inferences
    are possible, only WD 0718+3731 has a statistically significant
    color and flux excess consistent with a giant planet origin.}
    \label{tab:with_excess}
\end{table}

We report in Table \ref{tab:with_excess} the white dwarfs in our sample
with statistically significant color and flux excesses.  We also provide a
complete list of the 51 white dwarfs in our sample in the Appendix.  If we
further require both (1) no statistically significant [3.6] flux excess
and (2) an absolute [3.6] excess less than 0.5 mag, then two candidates
remain: WD 0003+4336 and WD 0718+3731 with color excesses significant
at the 2.1$\sigma$ and 4.3$\sigma$ levels.  At our inferred total ages
for these white dwarfs, their observed excesses could be produced by
giant planets with $M_{\text{p}} = 4.9~M_{\text{Jup}}$ and $M_{\text{p}}
= 3.6~M_{\text{Jup}}$.  However, we expect that in our sample of 51 white
dwarfs one will have a color excess significant at the 2$\sigma$ level due
to chance alone.  Indeed, we also find one object with a color deficiency
(rather than excess) significant at the 2.4$\sigma$ level, confirming the
fidelity of our uncertainty estimates.  The probability of identifying
a white dwarf with a color excess significant at the 4.3$\sigma$ level
due to randomness is extremely low (0.001 for a sample of size 51).  As a
result, we conclude that the color excess we observe in the WD 0718+3731
system is astrophysical in origin and best explained by the presence of
an unresolved giant planet.  On the other hand, we attribute the color
excess we observe in the WD 0003+4336 system to randomness.  We give in
Table \ref{tab:candidate} the properties of the candidate giant planet
in the WD 0718+3731 system if its color and flux excesses are interpreted
as a consequence of the presence of an unresolved giant planet companion.

Our color and flux excess criteria give us sensitivity to objects with
$M \gtrsim 6~M_{\text{Jup}}$ for white dwarfs with $M_{\text{WD}}
\approx 0.7~M_{\odot}$ and sensitivity to objects with $M \gtrsim
2~M_{\text{Jup}}$ for white dwarfs with $M_{\text{WD}} \approx
1.4~M_{\odot}$.  Our requirement that the absolute [3.6] flux excess
be less than 0.5 mag allows us to differentiate between unresolved dust
disks and giant planet-mass objects at the cost of sensitivity to brown
dwarfs.  For white dwarfs with $M_{\text{WD}} \approx 0.7~M_{\odot}$,
we are insensitive to brown dwarfs with $M \gtrsim 50~M_{\text{Jup}}$.
For white dwarfs with $M_{\text{WD}} \approx 1.4~M_{\odot}$, we are
insensitive to brown dwarfs with $M \gtrsim 10~M_{\text{Jup}}$.

\section{A Candidate Giant Planet in the GALEX J071816.4+373139
System}\label{sec:cand}

The white dwarf WD 0718+3731 referred to by its Simbad designation GALEX
J071816.4+373139 from here has a [3.6]--[4.5] color excess significant at
the $4.3\sigma$ level, a [4.5] flux excess significant at the $3.1\sigma$
level, and no significant [3.6], [5.8], or [8.0] flux excesses.
We emphasize that the significances of these excesses are independent
of any white dwarf model.  If we use the total age for the white dwarf
$\tau = 0.2$ Gyr inferred using the methodology described in Section
\ref{sec:data}, a white dwarf model atmosphere with its $T_{\text{eff}}$,
and the cloudless models for giant planets and brown dwarfs published
by \citet{Marley_2021}, then the [3.6]--[4.5] color excess we observe
for GALEX J071816.4+373139 can be used to infer the mass of a potential
unresolved giant planet-mass object responsible for the color excess.
We find that all of our observations and inferences for the GALEX
J071816.4+373139 system can be self-consistently explained by the presence
of an unresolved giant planet-mass object with $M_{\text{p}} \approx
3.6~M_{\text{Jup}}$ and the properties given in Table \ref{tab:candidate}.

\begin{table*}
    \centering
    \begin{tabular}{cccc|cc|ccc|ccc}
    \hline
    White dwarf & Gaia DR3 & $d$ & $\tau$ & $M_\text{WD}$ & $M_{\ast,i}$ & $M_{\text{p}}$ & $T_{\text{eff,p}}$ & $\log{g}$ & [3.6] flux & [4.5] flux & [4.5] flux excess \\
    & \texttt{source\_id} & (pc) & (Gyr) & ($M_{\odot}$) & ($M_{\odot}$) & ($M_\text{Jup}$) & (K) & & ($\mu$Jy) & ($\mu$Jy) & ($\mu$Jy) \\
    \hline
    0718+3731 & 898348313253395968 & 85 & 0.2  & 1.37 & 8 & 3.6 & 400 & 3.8 & 16.5$\pm$0.6 & 15.5$\pm$0.7 & 4.9$\pm$0.8 \\
    \hline
    \end{tabular}
    \caption{Properties of our candidate white dwarf--unresolved giant
    planet system}
    \label{tab:candidate}
\end{table*}

It is possible that the [4.5] flux excess we observe in the GALEX
J071816.4+373139 system could be explained by a dust disk.  A disk of
dust with $T_{\text{dust}} \approx 650$ K could produce a [4.5] flux
excess, but it would be unlikely to produce the color excess we observe.
In any case, such a dust disk would produce significant [5.8] and [8.0]
flux excesses that we do not observe.  We therefore argue that a dust
disk with $T_{\text{dust}} \approx 650$ K is inconsistent with our
observations of the GALEX J071816.4+373139 system.

An unresolved optically faint but mid-infrared bright background
high-redshift star-forming galaxy with strong emission lines could
conceivably produce the [3.6]-[4.5] color excess, the [4.5] flux
excess, and nondetections of [3.6], [5.8], and [8.0] flux excesses
we observe in the GALEX J071816.4+373139 system.  To investigate the
likelihood of this scenario, we use archival Spitzer/IRAC data in the
COSMOS field \citep{Sanders_2007} to quantify the density on sky of
background galaxies that could create a similar observational signature
to what we observe in the GALEX J071816.4+373139 system.  We find that
a chance alignment with a background galaxy that could create a similar
color excess has only a 0.005 (i.e., 1 in 200) chance of producing an
event like that we observe in the GALEX J071816.4+373139 system in our
51-star sample.  We note that GALEX J071816.4+373139 lies in projection
about $1^{\circ}$ from the core of the galaxy cluster MACS J0717.5+3745.
Indeed, the Spitzer data we use in our analysis were collected by Program
40652 studying that cluster.  The cluster has $z = 0.546$, so any star
forming galaxies with strong emission lines in that cluster would be
visible in existing optical and near-infrared data at the location of
GALEX J071816.4+373139.  We therefore argue that an optically faint but
mid-infrared bright background high-redshift star forming galaxy with
strong emission lines is unlikely to explain all of our observations
for the GALEX J071816.4+373139 system.

GALEX J071816.4+373139 is very massive: $M_{\text{WD}} = 1.27~M_{\odot}$
assuming the oxygen--neon white dwarf model from \citet{Camisassa_2019}
or $M_{\text{WD}} = 1.29~M_{\odot}$ assuming the carbon--oxygen
white dwarf model from \citet{Bedard_2020}.  GALEX J071816.4+373139 is
therefore among the most massive white dwarfs in the solar neighborhood
\citep{Kilic_2021_massive}.  Assuming GALEX J071816.4+373139 was the
result of single stellar evolution, then its progenitor had an initial
mass in the range $6~M_{\odot} \lesssim M_{\ast,i} \lesssim 10~M_{\odot}$
with the value dependent on the specific \citet{Cummings_2018}
initial--final mass relation adopted.  As a consequence, the giant planet
candidate in the GALEX J071816.4+373139 system would have formed around
the most massive star known to have hosted a planet while it was on the
main sequence.

\begin{figure*}
    \centering
    \includegraphics[width=\textwidth]{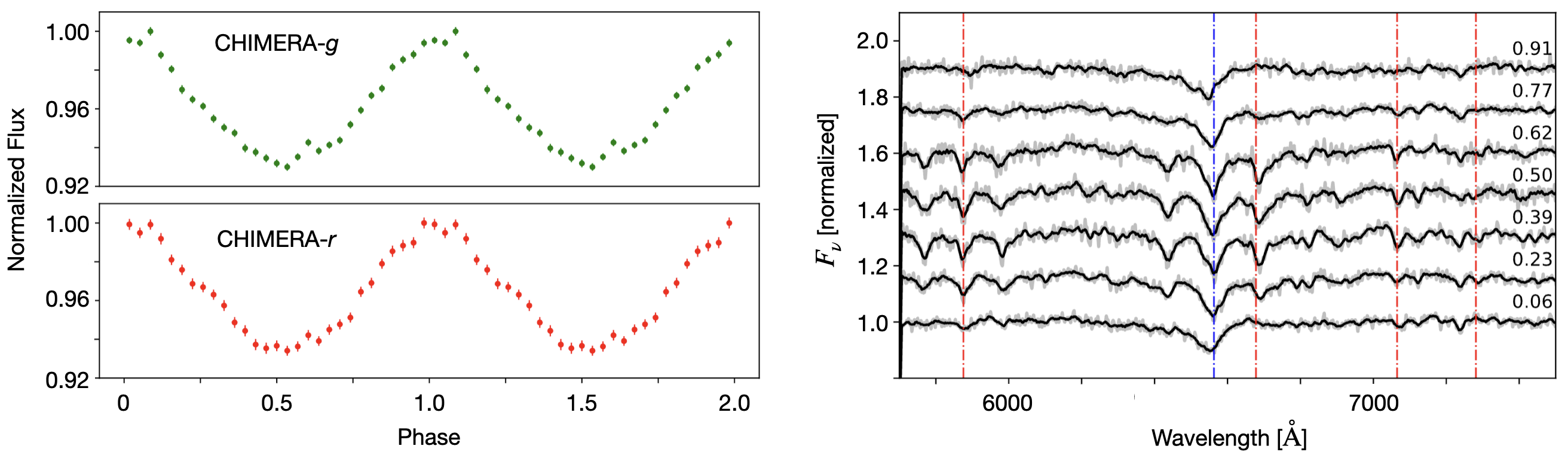}
    \caption{Time-resolved photometric and spectroscopic observations of
    GALEX J071816.4+373139.  Top left: phase-folded $g'$ light curve
    from the high-speed imaging photometer CHIMERA on the 5m Hale
    Telescope \citep{Harding_2016}.  Bottom left: phase-folded $r'$
    CHIMERA light curve.  GALEX J071816.4+373139 varies by about 6\%
    peak-to-peak with a period of 11.3 minutes.  Right: phase-resolved
    Keck/LRIS spectra of GALEX J071816.4+373139 normalized and shifted
    vertically for clarity with the numbers on the right side of the
    plot indicating the center of each phase bin.  Phase 0 corresponds
    to the maximum in the light curve illustrated on the left.  The
    blue vertical line highlights hydrogen H$\alpha$ while the red
    lines indicate absorption lines of neutral helium.  At phase 0.5,
    H$\alpha$ and the helium 5875~\AA~line are almost equally strong, and
    their Zeeman splitting implies a magnetic field with $|\mathbf{B}|
    \approx 8$ MG.  At phase 0, only hydrogen is present and the absence
    of splitting indicates $|\mathbf{B}| \lesssim 1$ MG.}
    \label{fig:T1}
\end{figure*}

Our mass inference for the giant planet-mass object in the GALEX
J071816.4+373139 system relies on our inferred total age for GALEX
J071816.4+373139, which assumes that the white dwarf was the result of
single stellar evolution.  If the white dwarf GALEX J071816.4+373139 was
formed through the merger of two lower-mass white dwarfs, then those
white dwarfs would have had much longer main-sequence lifetimes.  The
total age for the system would then be much longer than we inferred,
and the mass of the unresolved object responsible for the color and flux
excesses would be much more massive, almost certainly a brown dwarf.
In that case, because both the two now-merged white dwarfs would have had
main-sequence masses well below the mass of a main-sequence B stars,
the GALEX J071816.4+373139 system would not be useful to constrain
the occurrence of giant planets orbiting massive main-sequence stars.
GALEX J071816.4+373139 is known to be photometrically variable (left
panels of Figure \ref{fig:T1}), and this variability has been attributed
to the rotation of an inhomogeneous surface created by a strong magnetic
field sometimes predicted to be found in massive white dwarfs formed
via mergers.

If the massive white dwarf GALEX J071816.4+373139 was the result of a
merger between two lower-mass white dwarfs and therefore old, then it
would be expected to have ``hot'' Galactic kinematics observable as
a high tangential velocity \citep{Cheng_2020} in addition to a strong
magnetic field and fast rotation \citep{Kilic_2023}.  The tangential
velocity of GALEX J071816.4+373139 with respect to the local standard
of rest is about 10 km s$^{-1}$.  Assuming the \citet{Holmberg_2009}
age--velocity dispersion relation for the Milky Way's disk, the age
implied by this transverse velocity is consistent with the our single
stellar evolution-based age inference, thereby supporting our mass
inference for the unresolved object in the GALEX J071816.4+373139 system.
Likewise, while GALEX J071816.4+373139 is thought to have a strong
magnetic field, it is still debated whether all strong magnetized and
fast rotating white dwarfs are produced by mergers \citep{Ferrario_2015}.

In addition to its photometric variability, GALEX J071816.4+373139 is
spectroscopically variable as well.  Time-resolved Keck/Low-Resolution
Imaging Spectrometer (LRIS) data have revealed that GALEX J071816.4+373139
is ``two faced'', with one hemisphere hydrogen dominated and
the other helium dominated (right panel of Figure \ref{fig:T1}).
As argued by \citet{Caiazzo_2023}, this two-faced nature is likely
the result of inhomogeneous surface magnetic fields that result in
temperature, pressure, or mixing strength variation over the surface.
\citet{Caiazzo_2023} suggested that as GALEX J071816.4+373139 cools it is
undergoing a transition from a hydrogen-dominated to a helium-dominated
atmosphere.  The strength of GALEX J071816.4+373139's magnetic field
could generate cyclotron emission that could produce a [4.5] flux excess.
Like the dust disk scenario, the power-law nature of cyclotron emission
that produces a [4.5] flux excess would also likely produce [3.6],
[5.8], and [8.0] flux excesses that we do not observe.  Cyclotron
emission would also require ongoing accretion, and as described above,
our analysis rules out the presence of a brown-dwarf mass companion to
GALEX J071816.4+373139.  We therefore argue that power-law cyclotron
emission as a consequence of ongoing accretion is inconsistent with our
observations of the GALEX J071816.4+373139 system.

Our mass inferences for the unresolved companion in the GALEX
J071816.4+373139 system responsible for the [3.6]--[4.5] color and
[4.5] flux excesses we observe relies on the \citet{Marley_2021} grid
of evolution and atmosphere models for brown dwarfs and giant planets.
Those models assume cloudless atmospheres that are likely appropriate
for our inferred temperature of the unresolved companion in the GALEX
J071816.4+373139 system.  Nevertheless, if the unresolved companion in the
GALEX J071816.4+373139 system is cloudy, then its mass would be somewhat
larger than the $M_{\text{p}} \approx 3.6~M_{\text{Jup}}$ we infer based
on our observed color excess using the cloudless \citet{Marley_2021}
models.  In addition, non-equilibrium chemistry resulting from vertical
mixing may reduce the 4.5 micron flux of giant planet-mass objects by
about 50\% \citep[e.g.,][]{Phillips_2020}.  Our mass estimate for the
unresolved companion in the GALEX J071816.4+373139 system is therefore
likely a lower limit.

\section{The Occurrence of Giant Planets Orbiting B Stars}\label{sec:pop}

In addition to the identification of an unresolved giant planet candidate
in the GALEX J071816.4+373139 system, our non-detections of color excesses
attributable to unresolved giant planets or brown dwarfs in our sample
of young, massive white dwarfs also enable us to infer the occurrence
of giant planets orbiting stars with $M_{\ast,i} \gtrsim 3~M_{\odot}$.
In addition to our detection of a planet candidate with $M_{\text{p}}
\approx 3.6~M_{\text{Jup}}$, our occurrence inference depends on the
number of white dwarfs with sufficiently precise Spitzer/IRAC photometry
in the SEIP Source List to enable the detection of planet-mass objects.
We define our effective sample size $N_{\text{WD}}(M_{\text{p}})$ as a
function of planet mass as the number of white dwarfs for which a planet
of mass $M_{\text{p}}$ would produce a $2\sigma$ color excess detection.

Since some of the massive, apparently young white dwarfs in our sample
will be the result of mergers between two lower-mass white dwarfs
and therefore much older than suggested by our analyses that assumed
single stellar evolution, we must also correct $N_{\text{WD}}$ for this
merger fraction.  \citet{Cheng_2020} showed that about 25\% of massive
white dwarfs in the solar neighborhood are the result of mergers, so we
multiply $N_{\text{WD}}$ by 0.75 to account for these merger products.

\begin{table*}
    \centering
    \begin{tabular}{c|cccc}
    \hline
    Mass range & sample size & sample size & number of & occurrence \\
    & uncorrected for mergers & corrected for mergers & candidates & \\
    \hline
    $3~M_{\text{Jup}} < M_{\text{p}} < 10~M_{\text{Jup}}$ & 12 & 9 & 1 & $0.11_{-0.07}^{+0.13}$ \\
%     &  &  & \textbf{0 (if merger origin)} & $<\textbf{0.024}$ \\
    $5~M_{\text{Jup}} < M_{\text{p}} < 10~M_{\text{Jup}}$ & 30 & 23 & 0 & $<0.0099$ \\
    $10~M_{\text{Jup}} < M_{\text{p}} < 20~M_{\text{Jup}}$ & 29 & 22 & 0 & $<0.010$ \\
    $10~M_{\text{Jup}} < M_{\text{p}} < 30~M_{\text{Jup}}$ & 26 & 20 & 0 & $<0.011$ \\
    \hline
    \end{tabular}
    \caption{Occurrence as a function of candidate giant planet/brown
    dwarf mass.  We note that if GALEX J071816.4+373139 did form as
    the result of a merger between two lower-mass white dwarfs, then
    our inference of the occurrence of objects with $3~M_{\text{Jup}}
    < M_{\text{p}} < 10~M_{\text{Jup}}$ orbiting main-sequence B stars
    would be $\eta_{\text{GP}} < 0.025$ (i.e., an upper limit with the
    value 0.025).}
    \label{tab:occurrence}
\end{table*}

\begin{figure*}
    \centering
    \includegraphics[width=0.7\textwidth]{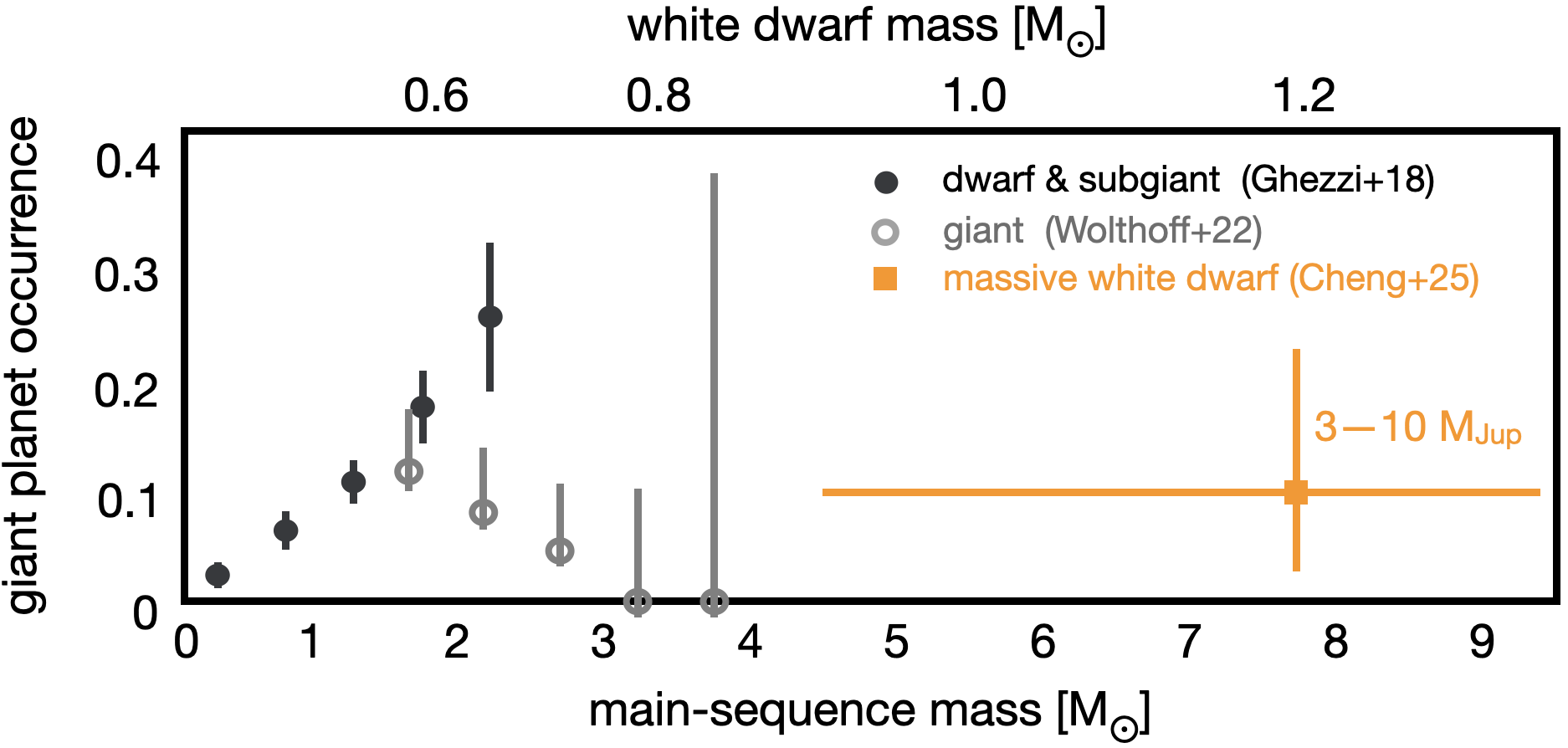}
    \caption{Dependence of giant planet occurrence on host star mass.}
    \label{fig:occurrence}
\end{figure*}

We model the number of giant planet/brown dwarf candidates
$N_{\text{cand}}$ in a sample of size $N_{\text{WD}}$ candidates using a
binomial distribution.  Following \citet{Schlaufman_2014}, we exploit the
fact that a Beta$(\alpha$,$\beta$) distribution is a conjugate prior to
the binomial distribution and will result in a Beta distribution posterior
for the occurrence of giant planet/brown dwarf candidates in a sample.
Bayes's theorem guarantees
\begin{eqnarray}
f(\theta|\mathbf{y}) & = & \frac{f(\mathbf{y}|\theta)f(\theta)}
                                {\int f(\mathbf{y}|\theta)f(\theta)d\theta},
\end{eqnarray} 
where $f(\theta|\mathbf{y})$ is the posterior distribution of the model
parameter $\theta$, $f(\mathbf{y}|\theta)$ is the likelihood of the data
$\mathbf{y}$ given $\theta$, and $f(\theta)$ is the prior for $\theta$.
In this case, the likelihood is the binomial likelihood that describes
the probability of a number of successes $y$ in $n$ Bernoulli trials
each with probability $\theta$ of success:
\begin{eqnarray}
f(y|\theta) = \left(\begin{array}{cc} n \\ y \end{array} \right)
              \theta^{y} \left(1-\theta\right)^{n-y}.
\end{eqnarray} 
As shown by \citet{Schlaufman_2014}, in this situation using a
Beta($\alpha$,$\beta$) prior on $\theta$ with hyperparameters $\alpha$
and $\beta$ results in a Beta posterior for $\theta$ of the form
Beta($\alpha+N_{\text{cand}},\beta+N_{\text{WD}}-N_{\text{cand}}$).

The hyperparameters $\alpha$ and $\beta$ of the prior can be thought
of as encoding a certain amount of prior information in the form of
pseudo-observations.  Specifically, $\alpha-1$ is the number of successes
and $\beta-1$ is the number of failures imagined to have already been
observed and therefore included as prior information on $\theta$.
Taking any $\alpha = \beta = i$ where $i \geq 1$ could be thought of as
an uninformative prior in the sense that the probabilities of success
and failure in the prior distribution are equally likely.  However, if
$i$ is large, then there is imagined to be a lot of prior information,
and the posterior distribution will mostly reflect the prior when $n
\leq i$.  On the other hand, if $n \gg i$, then the posterior will
be dominated by the data.  Since in our case both $N_{\text{cand}}$
and $N_{\text{WD}}$ are small, when $N_{\text{cand}} > 0$ we choose the
improper prior $\alpha = \beta = 0.3$ such that our posterior median is
equal to $N_{\text{cand}}/N_{\text{WD}}$.  We then use the same $\alpha
= \beta = 0.3$ when $N_{\text{cand}} = 0$ and quote as upper limits the
upper boundaries of credible intervals containing 68\% of the occurrence
posteriors.

\begin{figure*}
\centering
\includegraphics[width=0.8\textwidth]{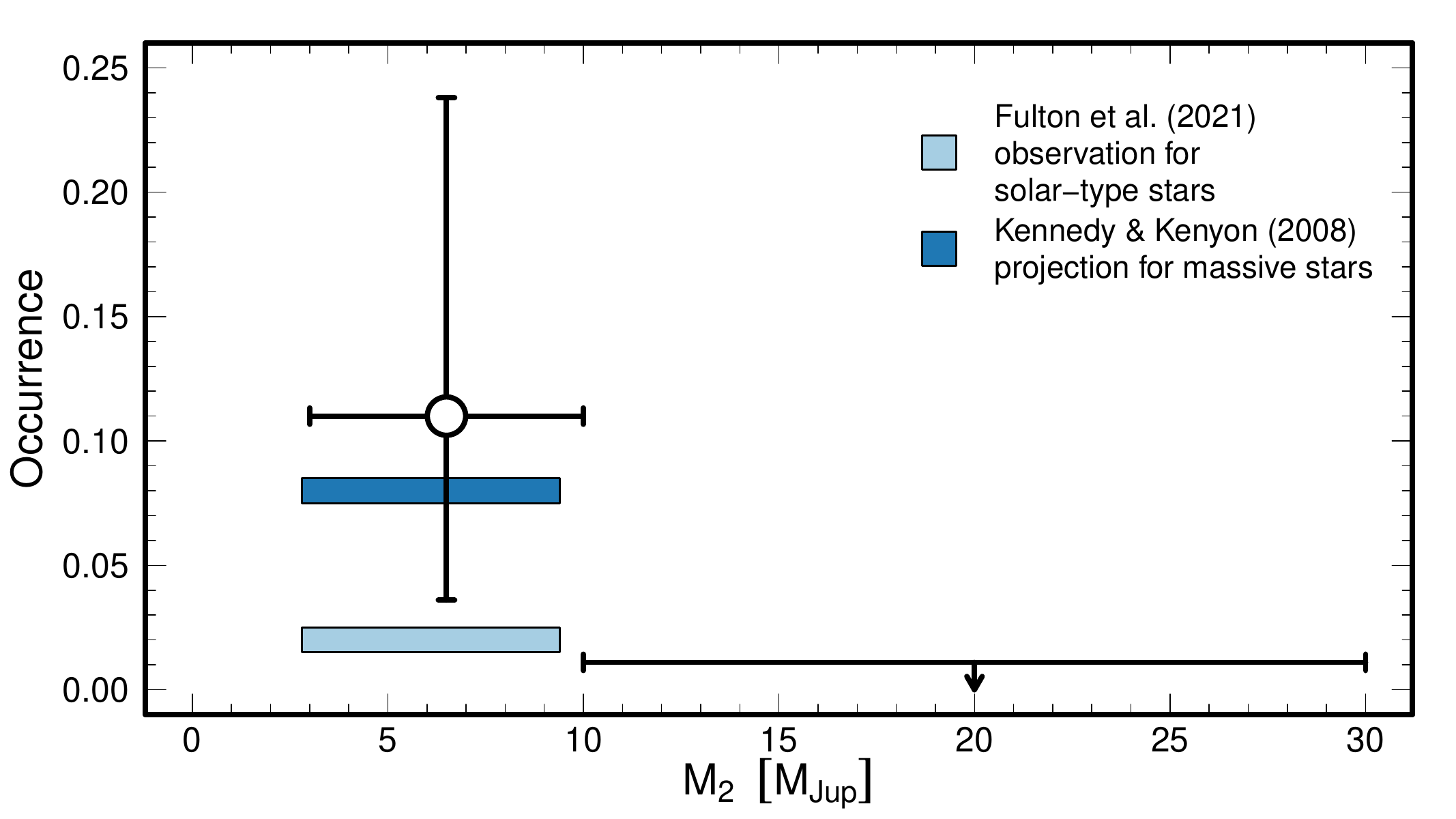}
\caption{Occurrence of giant planets and brown dwarfs as function of mass.
For stars with $M_{\ast,i} \gtrsim 3~M_{\odot}$, we find $\eta_{\text{GP}}
= 0.11_{-0.07}^{+0.13}$ in the mass range $3~M_{\text{Jup}} \lesssim
M_{\text{p}} \lesssim 10~M_{\text{Jup}}$, and $\eta_{\text{BD}} < 0.011$
in the mass range $10~M_{\text{Jup}} \lesssim M_{\text{p}} \lesssim
30~M_{\text{Jup}}$.  The former value is in accord with the factor of
four increase in $\eta_{\text{GP}}$ predicted by \citet{Kennedy_2008}
for stars with $3~M_{\odot} \lesssim M_{\ast} \lesssim 4~M_{\odot}$
relative to solar-mass dwarfs.\label{fig:occ_comp}}
\end{figure*}

We give in Table \ref{tab:occurrence} $N_{\text{WD}}(M_{\text{p}})$
both uncorrected for and corrected for mergers as well as our overall
occurrence inferences.  We plot in Figure~\ref{fig:occurrence} occurrence
as a function of host stellar mass converted from white dwarf mass
using the open cluster-based \citet{Cummings_2018} initial--final mass
relation.  To convert the planet occurrence around white dwarfs into
that around the progenitor stars, we account for planet survival and
orbital migration.  When a star evolves into a white dwarf, shedding
several solar masses into space, planetary orbits expand stably and
adiabatically by a factor proportional to the initial--final mass ratio
of the host star \citep{Veras_2011, Veras_2020}.  Planets with small
orbital separations will be tidally engulfed during the giant phase, but
planets on wider orbits are expected to survive.  \citet{Mustill_2012}
suggest minimum initial orbit radii $4~\text{au} \lesssim a \lesssim
7$ au and final orbit radii 8au$\lesssim a \lesssim$20au for
stars with $3~M_{\odot} \lesssim M_{\odot,i} \lesssim 9~M_{\odot}$
(i.e., $0.7~M_{\odot} \lesssim M_{\text{WD}} \lesssim 1.4~M_{\odot}$).
Therefore, the occurrence we constrain is approximately equal to the
occurrence of giant planets orbiting stars with $M_{\ast,i} \gtrsim
3~M_{\odot}$ with $a\gtrsim 5$au.

Theoretical predictions for the occurrence of giant planets orbiting
massive stars sensitively depend on the assumptions made in models of
planet formation for the relationship between host star mass, disk mass,
and disk lifetime \citep[e.g.,][]{Alibert_2011}.  \citet{Kennedy_2008}
predicted that giant planet occurrence should increase by about a factor
of four from the occurrence observed for solar-mass dwarf stars to stars
with $3~M_{\odot} \lesssim M_{\ast} \lesssim 4~M_{\odot}$.  In contrast,
studies of evolved stars with masses in the range $2~M_{\odot} \lesssim
M_{\ast} \lesssim 3~M_{\odot}$ seem to suggest that the occurrence
of Doppler-detectable giant planets orbiting evolved massive stars
peaks at $\eta_{\text{GP}} \approx 0.10$ for $M_{\ast} \approx
1.8~M_{\odot}$ and drops down to $\eta_{\text{GP}} \approx 0.02$
for $M_{\ast} \approx 3~M_{\odot}$ albeit with large uncertainties
\citep[e.g.,][]{Wolthoff_2022}.

We find $\eta_{\text{GP}} = 0.11_{-0.07}^{+0.13}$ in the
mass range $3~M_{\text{Jup}} \lesssim M_{\text{p}} \lesssim
10~M_{\text{Jup}}$ for stars with $M_{\ast,i} \gtrsim 3~M_{\odot}$.
\citet{Fulton_2021} report $\eta_{\text{GP}} \approx 0.02 \pm 0.005$
in the mass range $3~M_{\text{Jup}} \lesssim M_{\text{p}} \lesssim
10~M_{\text{Jup}}$ for solar-mass dwarf stars.  In accord with the
\citet{Kennedy_2008} prediction, as we plot in Figure \ref{fig:occ_comp}
our $\eta_{\text{GP}}$ inference for massive giant planets orbiting
stars with $M_{\ast,i} \gtrsim 3~M_{\odot}$ is consistent with a factor
of four increase in the \citet{Fulton_2021} value.  At the same time,
our $\eta_{\text{GP}}$ inference is also consistent with the value
reported by \citet{Wolthoff_2022}.  Future JWST NIRCam observations of approximately 250
additional white dwarfs from our sample of 2195 Gaia-identified massive,
young white dwarfs would provide the statistical precision necessary
evaluate whether or not giant planet occurrence takes on its maximum
value in the range $3~M_{\odot} \lesssim M_{\ast} \lesssim 4~M_{\odot}$.

\section{Conclusion}\label{sec:conc}

We used Spitzer/IRAC SEIP Source List photometry to search for unresolved
mid-infrared color excesses indicative of unresolved giant planet-mass
objects orbiting 51 massive, young white dwarfs with Spitzer/IRAC [3.6]
and [4.5] data selected from a larger sample of 2195 massive, young
solar-neighborhood white dwarfs.  We identify one candidate in the GALEX
J071816.4+373139 system with $M_{\text{p}} \approx 3.6~M_{\text{Jup}}$,
$T_{\text{eff}} \approx 400$ K, and $\log{g} \approx 3.8$.  We also use
these data to infer the occurrence of giant planets $\eta_{\text{GP}}
= 0.11_{-0.07}^{+0.13}$ in the mass range $3~M_{\text{Jup}} \lesssim
M_{\text{p}} \lesssim 10~M_{\text{Jup}}$ for stars with initial
masses $M_{\ast,i} \gtrsim 3~M_{\odot}$.  This $\eta_{\text{GP}}$
inference is consistent both with the predicted factor of 4 increase
in $\eta_{\text{GP}}$ for stars with $3~M_{\odot} \lesssim M_{\ast}
\lesssim 4~M_{\odot}$ relative to solar-mass dwarfs and with Doppler-based
$\eta_{\text{GP}}$ inferences for evolved massive stars.

While we argue that the Galactic kinematics of GALEX J071816.4+373139
support our inference that it is a product of the stellar evolution
of single B star and not the merger of two lower-mass white dwarfs, we
cannot conclusively eliminate the merger scenario from consideration.
If GALEX J071816.4+373139 is the result of the merger of two lower-mass
white dwarfs, then our analyses would suggest an upper limit on the
occurrence $\eta_{\text{GP}}< 0.025$ for objects with $3~M_{\text{Jup}}
\lesssim M_{\text{p}} \lesssim 10~M_{\text{Jup}}$ orbiting main-sequence
B stars.  Future JWST/NIRCam observations of additional massive, young
white dwarfs from our sample will be sensitive to Saturn-mass planets
and would provide the definitive measurement of $\eta_{\text{GP}}$
for stars with $M_{\ast,i} \gtrsim 3~M_{\odot}$.

% 6
\section*{Acknowledgments}

We thank Jay Farihi, Guangwei Fu, J.\ J.\ Hermes, Mary Anne Limbach,
and Daniel Thorngren for useful discussions.  S.C. thanks Siyu
Yao for her constant inspiration and encouragement.  S.C.
acknowledges the support of the Martin A.\ and Helen Chooljian
Member Fund, funding from the Zurich Insurance Company, and the
Fund for Natural Sciences at the Institute for Advanced Study.
This work has made use of data from the European Space Agency (ESA)
mission {\it Gaia} (\url{https://www.cosmos.esa.int/gaia}), processed
by the {\it Gaia} Data Processing and Analysis Consortium (DPAC,
\url{https://www.cosmos.esa.int/web/gaia/dpac/consortium}).  Funding for
the DPAC has been provided by national institutions, in particular the
institutions participating in the {\it Gaia} Multilateral Agreement.
This work is based in part on observations made with the Spitzer Space
Telescope, which is operated by the Jet Propulsion Laboratory, California
Institute of Technology under a contract with NASA.  This publication
makes use of data products from the Wide-field Infrared Survey Explorer,
which is a joint project of the University of California, Los Angeles, and
the Jet Propulsion Laboratory/California Institute of Technology, funded
by the National Aeronautics and Space Administration.  This research
has made use of the NASA Exoplanet Archive, which is operated by the
California Institute of Technology, under contract with the National
Aeronautics and Space Administration under the Exoplanet Exploration
Program.  This research has made use of NASA's Astrophysics Data System.

\vspace{5mm}
\facilities{ADS, ESO:VISTA, Exoplanet Archive, Gaia, IRSA, NEOWISE,
Spitzer, UKIRT, WISE}

\software{\texttt{astropy} \citep{astropy_1, astropy_2, astropy_3}, 
\texttt{numpy} \citep{harris2020array}, 
\texttt{matplotlib} \citep{Hunter_2007},
\texttt{R} \citep{R},
\texttt{SciPy} \citep{scipy}}

\appendix

In Table \ref{tab:sample} below we provide a complete list of the 51
white dwarfs selected from our input sample of 2195 white dwarfs with
Spitzer/IRAC SEIP Source List [3.6] and [4.5] photometry.  We also
include for each white dwarf our inferred $T_{\text{eff}}$, $\log{g}$,
mass, age, color excess, flux excess, and mass range in which a companion
would be detectable.  We calculate these detectability bounds using the
total age of each white dwarf and the uncertainty of its SEIP photometry.
We consider a companion detectable when it would produce a color excess
in the system 3 times the SEIP color uncertainty for the system.
As companion mass increases through the giant planet/brown dwarf regime,
flux excesses in both [3.6] and [4.5] increase.  The color excess between
those two bands is expected to first increase and then decrease, though,
as a very massive companion would have a hot temperature and leave the
color index close to the Rayleigh-Jeans value.  Consequently, our color
excess approach is insensitive to both low-mass and high-mass companions.

\begin{longrotatetable}
\begin{deluxetable*}{lCCCCCCCCCCC}
    \tablecaption{51 Massive, Young White Dwarfs with Spitzer/IRAC
    SEIP Photometry}
    \startdata
    \tablehead{
    \colhead{Gaia DR3 \texttt{source\_id}} & \colhead{Distance} & \colhead{Envelope Type} & \colhead{$T_\text{eff}$} & \colhead{$\log{g}$} & \colhead{WD Mass} & \colhead{Age} & \colhead{$\Sigma$} & \colhead{$\chi_{3.6}$} & \colhead{$\chi_{4.5}$} & \multicolumn{2}{c}{Sensitivity Range} \\
   \colhead{} & \colhead{(pc)} & \colhead{} & \colhead{(K)} & \colhead{} & \colhead{($M_{\odot}$)} & \colhead{(Gyr)} & \colhead{} & \colhead{} & \colhead{} & \colhead{($M_{\text{Jup}}$)} & \colhead{($M_{\text{Jup}}$)}
    }
    384841336150015360  & 127.1 & \text{H } & 43260 &  8.98 & 1.204  &  0.073 & 2.085  & -1.09 & 1.30  & 3.59 &  11.6 \\
    292454841560140032  &  79.8 & \text{H } & 42561 &  9.07 & 1.243  &  0.094 & -0.74  & -0.36 & -0.94 & 2.60 &  11.6 \\
    95297185335797120   &  47.4 & \text{H } & 11666 &  8.26 & 0.770  &  1.049 & 22.44  & 52.68 & 120.6 & 4.67 &  41.3 \\
    94150944463779968   & 109.6 & \text{H } & 13269 &  8.25 & 0.767  &  0.869 &  1.52  & -1.36 & 0.53  & 14.5 &  41.6 \\
    5170668766392712448 &  70.3 & \text{H } & 32945 &  8.57 & 0.994  &  0.146 & -0.34  & -0.24 & -0.54 & 3.04 &  13.6 \\
    4613612951211823104 &  29.5 & \text{He} & 31805 &  9.25 & 1.327  &  0.388 & -0.18  & 0.29  & 0.17  & 1.59 &  13.4 \\
    4613612951211823616 &  29.5 & \text{H } & 16901 &  8.39 & 0.859  &  0.554 & -0.85  & -0.48 & -1.23 & 3.85 &  33.4 \\
    5060502958330979840 & 134.4 & \text{He} & 15609 &  8.31 & 0.786  &  0.707 &  0.38  & 0.51  & 0.71  & 10.5 &  36.8 \\
    216683956238279168  &  40.2 & \text{He} & 30047 &  8.63 & 1.027  &  0.168 & -1.52  & 0.78  & -0.74 & 6.58 &  40.1 \\
    3251244858154433536 &  31.2 & \text{H } & 45443 &  9.17 & 1.289  &  0.100 & -0.31  & 1.42  & 1.35  & 1.05 &  11.2 \\
    66697547870378368   & 130.1 & \text{He} & 35575 &  8.64 & 1.039  &  0.119 &  0.51  & -1.10 & -0.38 & 4.17 &  19.9 \\
    3302846072717868416 &  34.9 & \text{H } & 15018 &  8.31 & 0.806  &  0.694 &  0.03  & -0.37 & -0.43 & 4.37 &  37.9 \\
    45980377978968064   &  50.1 & \text{H } & 16673 &  8.39 & 0.861  &  0.564 &  0.22  & -0.89 & -0.85 & 4.36 &  34.7 \\
    3313606340183243136 &  44.9 & \text{He} & 24007 &  8.35 & 0.820  &  0.409 & -0.59  & 1.05  & 0.81  & 5.15 &  40.6 \\
    3313714023603261568 &  47.8 & \text{H } & 25626 &  8.17 & 0.739  &  0.536 &  0.00  & -0.39 & -0.46 & 6.51 &  45.0 \\
    147985404582267648  &  86.7 & \text{H } & 36564 &  8.63 & 1.035  &  0.114 & -0.78  & -1.02 & -1.52 & 9.57 &  32.5 \\
    3308403897837092992 &  46.0 & \text{He} & 16656 &  8.29 & 0.774  &  0.666 & -1.17  & -0.60 & -1.60 & 5.04 &  40.6 \\
    3201709827802584576 & 171.2 & \text{H } & 61895 &  8.82 & 1.143  &  0.053 &  1.46  & 0.26  & 1.71  & 11.3 &  11.8 \\
    % 254092090595748096  & 125.9 & \text{H } & 43403 &  6.72 & 0.277  &  nan   & -1.58  & -1.87 & -3.24 &\cdots&  \cdots \\
    % 4876967941937370368 & 137.7 & \text{H } &\cdots &\cdots & nan    &  nan   & -1.22  & -1.48 & -2.63 &\cdots&  \cdots \\
    3014049448078210304 &  83.5 & \text{H } & 36264 &  8.84 & 1.136  &  0.105 &  0.55  & -0.80 & -0.05 & 2.87 &  14.5 \\
    3216947242193857024 &  93.0 & \text{H } & 33628 &  8.75 & 1.092  &  0.123 & -0.37  & 0.27  & -0.08 & 2.44 &  16.2 \\
    4656950408184656128 &  81.5 & \text{He} & 16625 &  8.74 & 1.060  &  0.691 &  0.32  & 16.36 & 11.94 & 6.63 &  31.3 \\
    2884285498084601600 & 195.2 & \text{H } & 11092 &  7.80 & 0.743  &  0.494 &  0.18  & -0.55 & -0.54 & >80  &  >80 \\
    898348313253395968  &  85.3 & \text{He} & 37889 &  9.33 & 1.372  &  0.622 &  4.26  & -1.08 & 3.12  & 2.45 &  11.7 \\
    976040702520790400  &  32.4 & \text{H } & 14409 &  8.61 & 0.994  &  0.736 & -0.37  & -0.12 & -0.44 & 3.47 &  32.6 \\
    604972428842238080  &  77.6 & \text{H } & 11246 &  8.47 & 0.903  &  1.161 &  1.68  & -1.98 & -0.35 & 6.57 &  40.2 \\
    1044784918268692224 &  73.0 & \text{H } & 19579 &  8.16 & 0.720  &  0.675 &  1.51  & -0.83 & 0.17  & 6.43 &  44.2 \\
    % 5679722304793132800 & 120.3 & \text{H } &\cdots &\cdots & nan    &  nan   &  1.07  & 0.63  & 1.60  &\cdots& \cdots \\
    5461296765091298048 &  90.9 & \text{H } & 35304 &  8.97 & 1.192  &  0.134 &  0.92  & -0.43 & 0.66  & 3.23 &  13.2 \\
    % 850146827299032704  & 118.5 & \text{H } & 46252 &  7.89 & 0.634  &  1.271 & -1.04  & 1.12  & 0.59  & 18.6 &  72.3 \\
    834234385783616640  &  78.7 & \text{H } & 20275 &  8.14 & 0.707  &  0.720 & -2.41  & 1.58  & -0.19 & 6.12 &  41.8 \\
    1055273537642488576 &  62.6 & \text{H } & 24223 &  8.46 & 0.912  &  0.287 &  0.58  & -0.22 & 0.20  & 2.92 &  24.6 \\
    730459763934332416  &  37.8 & \text{H } & 22763 &  8.39 & 0.870  &  0.372 &  0.98  & -1.20 & -0.69 & 3.81 &  31.2 \\
    % 1074013923063975680 & 126.8 & \text{H } & 44984 &  7.93 & 0.651  &  1.092 &  0.13  & 0.46  & 0.67  & 32.9 &  75.7 \\
    3564376149017654272 &  46.8 & \text{H } & 24638 &  8.69 & 1.052  &  0.240 &  0.49  & -1.15 & -0.69 & 2.68 &  20.9 \\
    789339608048607232  &  48.2 & \text{H } & 12514 &  8.26 & 0.769  &  0.942 &  0.77  & -0.93 & -0.48 & 5.29 &  41.3 \\
    783880704600816000  &  85.1 & \text{He} & 34725 &  8.49 & 0.948  &  0.164 & -1.36  & 0.70  & -0.78 & 5.14 &  25.1 \\
    3698872156539379968 &  59.0 & \text{H } & 19670 &  8.30 & 0.811  &  0.497 &  0.69  & -0.79 & -0.39 & 4.34 &  33.9 \\
    1582231699483080192 & 199.4 & \text{H } & 12785 &  7.51 & 0.730  &  \cdots   & -1.35  & -1.15 & -2.25 & \cdots &  \cdots \\
    1572889389701360384 & 105.7 & \text{H } & 59018 &  7.99 & 0.708  &  0.629 &  0.55  & 0.32  & 0.80  & 16.0 &  70.4 \\
    % 3583181371265430656 &  32.7 & \text{He} & 11587 &  8.18 & 0.700  &  1.262 &  2.13  & 2.43  & 4.77  & 5.49 &  46.0 \\
    % 3904415787947492096 & 126.7 & \text{H } & 20168 &  8.06 & 0.661  &  1.056 & 21.86  & 82.6  & 167.2 & 8.08 &  55.3 \\
    1579147088331814144 &  59.3 & \text{He} & 28947 &  8.22 & 0.746  &  0.504 &  0.48  & 1.16  & 1.79  & 4.98 &  37.1 \\
    1551062778220116992 &  32.9 & \text{H } & 14610 &  8.37 & 0.844  &  0.699 &  0.31  & -0.64 & -0.52 & 4.13 &  34.8 \\
    1685171001732519808 &  85.3 & \text{He} & 36060 &  8.40 & 0.899  &  0.216 & -1.22  & -0.75 & -1.71 & 4.30 &  30.5 \\
    1479511096971183872 & 105.1 & \text{H } & 14150 &  8.19 & 0.731  &  0.862 & -1.05  & 0.85  & -0.25 & 7.43 &  41.7 \\
    1281410781322153216 &  48.3 & \text{H } & 22637 &  8.37 & 0.855  &  0.387 & -0.28  & 0.33  & 0.16  & 3.48 &  30.2 \\
    5779908502946006784 &  124. & \text{H } & 47344 &  9.08 & 1.246  &  0.066 & -1.04  & 3.84  & 1.53  & 2.03 &  11.4 \\
    6009760034351291904 &  89.8 & \text{H } & 51678 &  8.95 & 1.195  &  0.045 & -0.31  & 1.41  & 1.08  & 1.64 &  11.6 \\
    1202826348825240832 &  36.8 & \text{H } & 14788 &  8.40 & 0.868  &  0.681 & -0.21  & -0.51 & -0.75 & 4.33 &  34.7 \\
    1629292579563565696 & 137.5 & \text{He} & 38312 &  8.73 & 1.084  &  0.093 & -0.99  & -0.24 & -1.11 & 3.19 &  13.0 \\
    1356434445415005184 & 134.3 & \text{H } & 28175 &  8.28 & 0.811  &  0.373 & -0.84  & 0.03  & -0.73 & 6.65 &  33.4 \\
    1358301480583401728 &  31.5 & \text{H } & 30054 &  9.25 & 1.324  &  0.453 & -0.24  & -0.55 & -0.80 & 1.80 &  14.8 \\
    1634314740658456320 & 173.2 & \text{H } & 89534 &  7.90 & 0.732  &  0.527 & -0.47  & 0.43  & 0.18  & 27.9 &  79.3 \\
    5810075116283522688 & 130.1 & \text{H } & 46769 &  8.87 & 1.157  &  0.056 & -1.01  & 0.66  & -0.55 & 2.65 &  11.9 \\
    4078868662206175744 &  50.4 & \text{He} & 39161 &  8.28 & 0.830  &  0.321 & -0.18  & 2.12  & 2.31  & 4.63 &  33.6 \\
    2262849634963004416 &  12.8 & \text{He} & 12086 &  8.58 & 0.957  &  1.154 &  0.94  & 0.40  & 1.20  & 3.87 &  36.2 \\
    % 6461145119869641728 &  126. & \text{H } & 50506 &  7.91 & 0.655  &  1.044 &  0.29  & 1.65  & 2.23  & 15.4 &  70.1 \\
    2286107295188538240 &  71.3 & \text{He} & 13935 &  8.19 & 0.709  &  0.978 &  1.88  & 2.59  & 4.71  & 5.93 &  44.1 \\
    % 2810585920868186240 &  76.3 & \text{H } & 61771 &  7.83 & 0.645  &  1.154 & -0.08  & 0.11  & 0.07  & 61.9 &  >80 \\
    % 6499376994593401344 & 152.2 & \text{H } & 44442 &  7.94 & 0.655  &  1.042 &  0.78  & -0.25 & 0.43  & 26.0 &  71.3 \\
    % 6528109879126984960 & 114.0 & \text{H } & 34450 &  7.14 & 0.380  &  164.7 &  0.57  & 1.09  & 1.63  & \cdots &  \cdots 
    \enddata
\label{tab:sample}
\end{deluxetable*}
\end{longrotatetable}

% 7
\bibliography{article3}

\begin{thebibliography}{}
\expandafter\ifx\csname natexlab\endcsname\relax\def\natexlab#1{#1}\fi
\providecommand{\url}[1]{\href{#1}{#1}}
\providecommand{\dodoi}[1]{}
\providecommand{\doeprint}[1]{\href{http://ascl.net/#1}{\nolinkurl{http://ascl.net/#1}}}
\providecommand{\doarXiv}[1]{\href{https://arxiv.org/abs/#1}{\nolinkurl{https://arxiv.org/abs/#1}}}

\bibitem[{{Akeson} {et~al.}(2013){Akeson}, {Chen}, {Ciardi}, {Crane}, {Good},
  {Harbut}, {Jackson}, {Kane}, {Laity}, {Leifer}, {Lynn}, {McElroy}, {Papin},
  {Plavchan}, {Ram{\'\i}rez}, {Rey}, {von Braun}, {Wittman}, {Abajian}, {Ali},
  {Beichman}, {Beekley}, {Berriman}, {Berukoff}, {Bryden}, {Chan}, {Groom},
  {Lau}, {Payne}, {Regelson}, {Saucedo}, {Schmitz}, {Stauffer}, {Wyatt}, \&
  {Zhang}}]{Akeson_2013}
{Akeson}, R.~L., {Chen}, X., {Ciardi}, D., {et~al.} 2013,
  \href{http://dx.doi.org/10.1086/672273}{\color{magenta}\pasp},
  \href{https://ui.adsabs.harvard.edu/abs/2013PASP..125..989A}{\color{cyan}125},
  989

\bibitem[{{Alibert} {et~al.}(2011){Alibert}, {Mordasini}, \&
  {Benz}}]{Alibert_2011}
{Alibert}, Y., {Mordasini}, C., \& {Benz}, W. 2011,
  \href{http://dx.doi.org/10.1051/0004-6361/201014760}{\color{magenta}\aap},
  \href{https://ui.adsabs.harvard.edu/abs/2011A&A...526A..63A}{\color{cyan}526},
  A63

\bibitem[{{Astropy Collaboration} {et~al.}(2013){Astropy Collaboration},
  {Robitaille}, {Tollerud}, {Greenfield}, {Droettboom}, {Bray}, {Aldcroft},
  {Davis}, {Ginsburg}, {Price-Whelan}, {Kerzendorf}, {Conley}, {Crighton},
  {Barbary}, {Muna}, {Ferguson}, {Grollier}, {Parikh}, {Nair}, {Unther},
  {Deil}, {Woillez}, {Conseil}, {Kramer}, {Turner}, {Singer}, {Fox}, {Weaver},
  {Zabalza}, {Edwards}, {Azalee Bostroem}, {Burke}, {Casey}, {Crawford},
  {Dencheva}, {Ely}, {Jenness}, {Labrie}, {Lim}, {Pierfederici}, {Pontzen},
  {Ptak}, {Refsdal}, {Servillat}, \& {Streicher}}]{astropy_1}
{Astropy Collaboration}, {Robitaille}, T.~P., {Tollerud}, E.~J., {et~al.} 2013,
  \href{http://dx.doi.org/10.1051/0004-6361/201322068}{\color{magenta}\aap},
  \href{https://ui.adsabs.harvard.edu/abs/2013A&A...558A..33A}{\color{cyan}558},
  A33

\bibitem[{{Astropy Collaboration} {et~al.}(2018){Astropy Collaboration},
  {Price-Whelan}, {Sip{\H{o}}cz}, {G{\"u}nther}, {Lim}, {Crawford}, {Conseil},
  {Shupe}, {Craig}, {Dencheva}, {Ginsburg}, {VanderPlas}, {Bradley},
  {P{\'e}rez-Su{\'a}rez}, {de Val-Borro}, {Aldcroft}, {Cruz}, {Robitaille},
  {Tollerud}, {Ardelean}, {Babej}, {Bach}, {Bachetti}, {Bakanov}, {Bamford},
  {Barentsen}, {Barmby}, {Baumbach}, {Berry}, {Biscani}, {Boquien}, {Bostroem},
  {Bouma}, {Brammer}, {Bray}, {Breytenbach}, {Buddelmeijer}, {Burke},
  {Calderone}, {Cano Rodr{\'\i}guez}, {Cara}, {Cardoso}, {Cheedella}, {Copin},
  {Corrales}, {Crichton}, {D'Avella}, {Deil}, {Depagne}, {Dietrich}, {Donath},
  {Droettboom}, {Earl}, {Erben}, {Fabbro}, {Ferreira}, {Finethy}, {Fox},
  {Garrison}, {Gibbons}, {Goldstein}, {Gommers}, {Greco}, {Greenfield},
  {Groener}, {Grollier}, {Hagen}, {Hirst}, {Homeier}, {Horton}, {Hosseinzadeh},
  {Hu}, {Hunkeler}, {Ivezi{\'c}}, {Jain}, {Jenness}, {Kanarek}, {Kendrew},
  {Kern}, {Kerzendorf}, {Khvalko}, {King}, {Kirkby}, {Kulkarni}, {Kumar},
  {Lee}, {Lenz}, {Littlefair}, {Ma}, {Macleod}, {Mastropietro}, {McCully},
  {Montagnac}, {Morris}, {Mueller}, {Mumford}, {Muna}, {Murphy}, {Nelson},
  {Nguyen}, {Ninan}, {N{\"o}the}, {Ogaz}, {Oh}, {Parejko}, {Parley}, {Pascual},
  {Patil}, {Patil}, {Plunkett}, {Prochaska}, {Rastogi}, {Reddy Janga},
  {Sabater}, {Sakurikar}, {Seifert}, {Sherbert}, {Sherwood-Taylor}, {Shih},
  {Sick}, {Silbiger}, {Singanamalla}, {Singer}, {Sladen}, {Sooley},
  {Sornarajah}, {Streicher}, {Teuben}, {Thomas}, {Tremblay}, {Turner},
  {Terr{\'o}n}, {van Kerkwijk}, {de la Vega}, {Watkins}, {Weaver}, {Whitmore},
  {Woillez}, {Zabalza}, \& {Astropy Contributors}}]{astropy_2}
{Astropy Collaboration}, {Price-Whelan}, A.~M., {Sip{\H{o}}cz}, B.~M., {et~al.}
  2018, \href{http://dx.doi.org/10.3847/1538-3881/aabc4f}{\color{magenta}\aj},
  \href{https://ui.adsabs.harvard.edu/abs/2018AJ....156..123A}{\color{cyan}156},
  123

\bibitem[{{Astropy Collaboration} {et~al.}(2022){Astropy Collaboration},
  {Price-Whelan}, {Lim}, {Earl}, {Starkman}, {Bradley}, {Shupe}, {Patil},
  {Corrales}, {Brasseur}, {N{\"o}the}, {Donath}, {Tollerud}, {Morris},
  {Ginsburg}, {Vaher}, {Weaver}, {Tocknell}, {Jamieson}, {van Kerkwijk},
  {Robitaille}, {Merry}, {Bachetti}, {G{\"u}nther}, {Aldcroft},
  {Alvarado-Montes}, {Archibald}, {B{\'o}di}, {Bapat}, {Barentsen},
  {Baz{\'a}n}, {Biswas}, {Boquien}, {Burke}, {Cara}, {Cara}, {Conroy},
  {Conseil}, {Craig}, {Cross}, {Cruz}, {D'Eugenio}, {Dencheva}, {Devillepoix},
  {Dietrich}, {Eigenbrot}, {Erben}, {Ferreira}, {Foreman-Mackey}, {Fox},
  {Freij}, {Garg}, {Geda}, {Glattly}, {Gondhalekar}, {Gordon}, {Grant},
  {Greenfield}, {Groener}, {Guest}, {Gurovich}, {Handberg}, {Hart},
  {Hatfield-Dodds}, {Homeier}, {Hosseinzadeh}, {Jenness}, {Jones}, {Joseph},
  {Kalmbach}, {Karamehmetoglu}, {Ka{\l}uszy{\'n}ski}, {Kelley}, {Kern},
  {Kerzendorf}, {Koch}, {Kulumani}, {Lee}, {Ly}, {Ma}, {MacBride}, {Maljaars},
  {Muna}, {Murphy}, {Norman}, {O'Steen}, {Oman}, {Pacifici}, {Pascual},
  {Pascual-Granado}, {Patil}, {Perren}, {Pickering}, {Rastogi}, {Roulston},
  {Ryan}, {Rykoff}, {Sabater}, {Sakurikar}, {Salgado}, {Sanghi}, {Saunders},
  {Savchenko}, {Schwardt}, {Seifert-Eckert}, {Shih}, {Jain}, {Shukla}, {Sick},
  {Simpson}, {Singanamalla}, {Singer}, {Singhal}, {Sinha}, {Sip{\H{o}}cz},
  {Spitler}, {Stansby}, {Streicher}, {{\v{S}}umak}, {Swinbank}, {Taranu},
  {Tewary}, {Tremblay}, {de Val-Borro}, {Van Kooten}, {Vasovi{\'c}}, {Verma},
  {de Miranda Cardoso}, {Williams}, {Wilson}, {Winkel}, {Wood-Vasey}, {Xue},
  {Yoachim}, {Zhang}, {Zonca}, \& {Astropy Project Contributors}}]{astropy_3}
{Astropy Collaboration}, {Price-Whelan}, A.~M., {Lim}, P.~L., {et~al.} 2022,
  \href{http://dx.doi.org/10.3847/1538-4357/ac7c74}{\color{magenta}\apj},
  \href{https://ui.adsabs.harvard.edu/abs/2022ApJ...935..167A}{\color{cyan}935},
  167

\bibitem[{{Bang} {et~al.}(2020){Bang}, {Lee}, {Perdelwitz}, {Jeong}, {Han},
  {Oh}, \& {Park}}]{Bang_2020}
{Bang}, T.-Y., {Lee}, B.-C., {Perdelwitz}, V., {et~al.} 2020,
  \href{http://dx.doi.org/10.1051/0004-6361/201936613}{\color{magenta}\aap},
  \href{https://ui.adsabs.harvard.edu/abs/2020A&A...638A.148B}{\color{cyan}638},
  A148

\bibitem[{{Barber} {et~al.}(2016){Barber}, {Belardi}, {Kilic}, \&
  {Gianninas}}]{Barber_2016}
{Barber}, S.~D., {Belardi}, C., {Kilic}, M., \& {Gianninas}, A. 2016,
  \href{http://dx.doi.org/10.1093/mnras/stw683}{\color{magenta}\mnras},
  \href{https://ui.adsabs.harvard.edu/abs/2016MNRAS.459.1415B}{\color{cyan}459},
  1415

\bibitem[{{B{\'e}dard} {et~al.}(2020){B{\'e}dard}, {Bergeron}, {Brassard}, \&
  {Fontaine}}]{Bedard_2020}
{B{\'e}dard}, A., {Bergeron}, P., {Brassard}, P., \& {Fontaine}, G. 2020,
  \href{http://dx.doi.org/10.3847/1538-4357/abafbe}{\color{magenta}\apj},
  \href{https://ui.adsabs.harvard.edu/abs/2020ApJ...901...93B}{\color{cyan}901},
  93

\bibitem[{{Bergeron} {et~al.}(2011){Bergeron}, {Wesemael}, {Dufour},
  {Beauchamp}, {Hunter}, {Saffer}, {Gianninas}, {Ruiz}, {Limoges}, {Dufour},
  {Fontaine}, \& {Liebert}}]{Bergeron_2011}
{Bergeron}, P., {Wesemael}, F., {Dufour}, P., {et~al.} 2011,
  \href{http://dx.doi.org/10.1088/0004-637X/737/1/28}{\color{magenta}\apj},
  \href{https://ui.adsabs.harvard.edu/abs/2011ApJ...737...28B}{\color{cyan}737},
  28

\bibitem[{{Blouin} {et~al.}(2018){Blouin}, {Dufour}, \& {Allard}}]{Blouin_2018}
{Blouin}, S., {Dufour}, P., \& {Allard}, N.~F. 2018,
  \href{http://dx.doi.org/10.3847/1538-4357/aad4a9}{\color{magenta}\apj},
  \href{https://ui.adsabs.harvard.edu/abs/2018ApJ...863..184B}{\color{cyan}863},
  184

\bibitem[{{Brandner} {et~al.}(2021){Brandner}, {Zinnecker}, \&
  {Kopytova}}]{Brandner_2021}
{Brandner}, W., {Zinnecker}, H., \& {Kopytova}, T. 2021,
  \href{http://dx.doi.org/10.1093/mnras/staa3422}{\color{magenta}\mnras},
  \href{https://ui.adsabs.harvard.edu/abs/2021MNRAS.500.3920B}{\color{cyan}500},
  3920

\bibitem[{{Burleigh} {et~al.}(2002){Burleigh}, {Clarke}, \&
  {Hodgkin}}]{Burleigh_2002}
{Burleigh}, M.~R., {Clarke}, F.~J., \& {Hodgkin}, S.~T. 2002,
  \href{http://dx.doi.org/10.1046/j.1365-8711.2002.05417.x}{\color{magenta}\mnras},
  \href{https://ui.adsabs.harvard.edu/abs/2002MNRAS.331L..41B}{\color{cyan}331},
  L41

\bibitem[{{Burleigh} {et~al.}(2008){Burleigh}, {Clarke}, {Hogan}, {Brinkworth},
  {Bergeron}, {Dufour}, {Dobbie}, {Levan}, {Hodgkin}, {Hoard}, \&
  {Wachter}}]{Burleigh_2008}
{Burleigh}, M.~R., {Clarke}, F.~J., {Hogan}, E., {et~al.} 2008,
  \href{http://dx.doi.org/10.1111/j.1745-3933.2008.00446.x}{\color{magenta}\mnras},
  \href{https://ui.adsabs.harvard.edu/abs/2008MNRAS.386L...5B}{\color{cyan}386},
  L5

\bibitem[{{Burrows} {et~al.}(1997){Burrows}, {Marley}, {Hubbard}, {Lunine},
  {Guillot}, {Saumon}, {Freedman}, {Sudarsky}, \& {Sharp}}]{Burrows_1997}
{Burrows}, A., {Marley}, M., {Hubbard}, W.~B., {et~al.} 1997,
  \href{http://dx.doi.org/10.1086/305002}{\color{magenta}\apj},
  \href{https://ui.adsabs.harvard.edu/abs/1997ApJ...491..856B}{\color{cyan}491},
  856

\bibitem[{{Butler} {et~al.}(2006){Butler}, {Johnson}, {Marcy}, {Wright},
  {Vogt}, \& {Fischer}}]{Butler_2006}
{Butler}, R.~P., {Johnson}, J.~A., {Marcy}, G.~W., {et~al.} 2006,
  \href{http://dx.doi.org/10.1086/510500}{\color{magenta}\pasp},
  \href{https://ui.adsabs.harvard.edu/abs/2006PASP..118.1685B}{\color{cyan}118},
  1685

\bibitem[{{Caiazzo} {et~al.}(2023){Caiazzo}, {Burdge}, {Tremblay}, {Fuller},
  {Ferrario}, {G{\"a}nsicke}, {Hermes}, {Heyl}, {Kawka}, {Kulkarni}, {Marsh},
  {Mr{\'o}z}, {Prince}, {Richer}, {Rodriguez}, {van Roestel}, {Vanderbosch},
  {Vennes}, {Wickramasinghe}, {Dhillon}, {Littlefair}, {Munday}, {Pelisoli},
  {Perley}, {Bellm}, {Breedt}, {Brown}, {Dekany}, {Drake}, {Dyer}, {Graham},
  {Green}, {Laher}, {Kerry}, {Parsons}, {Riddle}, {Rusholme}, \&
  {Sahman}}]{Caiazzo_2023}
{Caiazzo}, I., {Burdge}, K.~B., {Tremblay}, P.-E., {et~al.} 2023,
  \href{http://dx.doi.org/10.1038/s41586-023-06171-9}{\color{magenta}\nat},
  \href{https://ui.adsabs.harvard.edu/abs/2023Natur.620...61C}{\color{cyan}620},
  61

\bibitem[{{Camisassa} {et~al.}(2019){Camisassa}, {Althaus}, {C{\'o}rsico}, {De
  Ger{\'o}nimo}, {Miller Bertolami}, {Novarino}, {Rohrmann}, {Wachlin}, \&
  {Garc{\'\i}a-Berro}}]{Camisassa_2019}
{Camisassa}, M.~E., {Althaus}, L.~G., {C{\'o}rsico}, A.~H., {et~al.} 2019,
  \href{http://dx.doi.org/10.1051/0004-6361/201833822}{\color{magenta}\aap},
  \href{https://ui.adsabs.harvard.edu/abs/2019A&A...625A..87C}{\color{cyan}625},
  A87

\bibitem[{{Carpenter} {et~al.}(2006){Carpenter}, {Mamajek}, {Hillenbrand}, \&
  {Meyer}}]{Carpenter_2006}
{Carpenter}, J.~M., {Mamajek}, E.~E., {Hillenbrand}, L.~A., \& {Meyer}, M.~R.
  2006, \href{http://dx.doi.org/10.1086/509121}{\color{magenta}\apjl},
  \href{https://ui.adsabs.harvard.edu/abs/2006ApJ...651L..49C}{\color{cyan}651},
  L49

\bibitem[{{Cheng}(2025)}]{WD_models}
{Cheng}, S. 2025,
  \href{http://dx.doi.org/10.5281/zenodo.15345675}{\color{magenta}zenodo},
  \href{https://zenodo.org/records/15345675}{\color{cyan}15345675}

\bibitem[{{Cheng} {et~al.}(2019){Cheng}, {Cummings}, \&
  {M{\'e}nard}}]{Cheng_2019}
{Cheng}, S., {Cummings}, J.~D., \& {M{\'e}nard}, B. 2019,
  \href{http://dx.doi.org/10.3847/1538-4357/ab4989}{\color{magenta}\apj},
  \href{https://ui.adsabs.harvard.edu/abs/2019ApJ...886..100C}{\color{cyan}886},
  100

\bibitem[{{Cheng} {et~al.}(2020){Cheng}, {Cummings}, {M{\'e}nard}, \&
  {Toonen}}]{Cheng_2020}
{Cheng}, S., {Cummings}, J.~D., {M{\'e}nard}, B., \& {Toonen}, S. 2020,
  \href{http://dx.doi.org/10.3847/1538-4357/ab733c}{\color{magenta}\apj},
  \href{https://ui.adsabs.harvard.edu/abs/2020ApJ...891..160C}{\color{cyan}891},
  160

\bibitem[{{Chilcote} {et~al.}(2022){Chilcote}, {Konopacky}, {Fitzsimmons},
  {Hamper}, {Macintosh}, {Marois}, {Savransky}, {Soummer}, {V{\'e}ran},
  {Agapito}, {Aleman}, {Ammons}, {Bonaglia}, {Boucher}, {Curliss}, {De Rosa},
  {Do {\'O}}, {Dunn}, {Esposito}, {Filion}, {Kerley}, {Landry}, {Lardiere},
  {Levinstein}, {Li}, {Limbach}, {Madurowicz}, {Maire}, {Millar-Blanchaer},
  {Nickson}, {Nielsen}, {Nguyen}, {Nguyen}, {Peng}, {Perera}, {Perrin}, {Por},
  {Poyneer}, {Pueyo}, {Rantakyr{\"o}}, {Sands}, {Spalding}, \&
  {Summey}}]{Chilcote_2022}
{Chilcote}, J., {Konopacky}, Q., {Fitzsimmons}, J., {et~al.} 2022, in
  \href{http://dx.doi.org/10.1117/12.2630159}{\color{magenta}Society of
  Photo-Optical Instrumentation Engineers (SPIE) Conference Series}, Vol.
  \href{https://ui.adsabs.harvard.edu/abs/2022SPIE12184E..1TC}{\color{cyan}12184},
  Ground-based and Airborne Instrumentation for Astronomy IX, ed. C.~J.
  {Evans}, J.~J. {Bryant}, \& K.~{Motohara}, 121841T

\bibitem[{{Cumming} {et~al.}(2008){Cumming}, {Butler}, {Marcy}, {Vogt},
  {Wright}, \& {Fischer}}]{Cumming_2008}
{Cumming}, A., {Butler}, R.~P., {Marcy}, G.~W., {et~al.} 2008,
  \href{http://dx.doi.org/10.1086/588487}{\color{magenta}\pasp},
  \href{https://ui.adsabs.harvard.edu/abs/2008PASP..120..531C}{\color{cyan}120},
  531

\bibitem[{{Cummings} {et~al.}(2018){Cummings}, {Kalirai}, {Tremblay},
  {Ramirez-Ruiz}, \& {Choi}}]{Cummings_2018}
{Cummings}, J.~D., {Kalirai}, J.~S., {Tremblay}, P.~E., {Ramirez-Ruiz}, E., \&
  {Choi}, J. 2018,
  \href{http://dx.doi.org/10.3847/1538-4357/aadfd6}{\color{magenta}\apj},
  \href{https://ui.adsabs.harvard.edu/abs/2018ApJ...866...21C}{\color{cyan}866},
  21

\bibitem[{{Dennihy} {et~al.}(2017){Dennihy}, {Clemens}, {Debes}, {Dunlap},
  {Kilkenny}, {O'Brien}, \& {Fuchs}}]{Dennihy_2017}
{Dennihy}, E., {Clemens}, J.~C., {Debes}, J.~H., {et~al.} 2017,
  \href{http://dx.doi.org/10.3847/1538-4357/aa8ef2}{\color{magenta}\apj},
  \href{https://ui.adsabs.harvard.edu/abs/2017ApJ...849...77D}{\color{cyan}849},
  77

\bibitem[{{Dye} {et~al.}(2018){Dye}, {Lawrence}, {Read}, {Fan}, {Kerr},
  {Varricatt}, {Furnell}, {Edge}, {Irwin}, {Hambly}, {Lucas}, {Almaini},
  {Chambers}, {Green}, {Hewett}, {Liu}, {McGreer}, {Best}, {Zhang}, {Sutorius},
  {Froebrich}, {Magnier}, {Hasinger}, {Lederer}, {Bold}, \& {Tedds}}]{Dye_2018}
{Dye}, S., {Lawrence}, A., {Read}, M.~A., {et~al.} 2018,
  \href{http://dx.doi.org/10.1093/mnras/stx2622}{\color{magenta}\mnras},
  \href{https://ui.adsabs.harvard.edu/abs/2018MNRAS.473.5113D}{\color{cyan}473},
  5113

\bibitem[{{Eisenhardt} {et~al.}(2020){Eisenhardt}, {Marocco}, {Fowler},
  {Meisner}, {Kirkpatrick}, {Garcia}, {Jarrett}, {Koontz}, {Marchese},
  {Stanford}, {Caselden}, {Cushing}, {Cutri}, {Faherty}, {Gelino}, {Gonzalez},
  {Mainzer}, {Mobasher}, {Schlegel}, {Stern}, {Teplitz}, \&
  {Wright}}]{Eisenhardt_2020}
{Eisenhardt}, P. R.~M., {Marocco}, F., {Fowler}, J.~W., {et~al.} 2020,
  \href{http://dx.doi.org/10.3847/1538-4365/ab7f2a}{\color{magenta}\apjs},
  \href{https://ui.adsabs.harvard.edu/abs/2020ApJS..247...69E}{\color{cyan}247},
  69

\bibitem[{{Fabricius} {et~al.}(2021){Fabricius}, {Luri}, {Arenou}, {Babusiaux},
  {Helmi}, {Muraveva}, {Reyl{\'e}}, {Spoto}, {Vallenari}, {Antoja}, {Balbinot},
  {Barache}, {Bauchet}, {Bragaglia}, {Busonero}, {Cantat-Gaudin}, {Carrasco},
  {Diakit{\'e}}, {Fabrizio}, {Figueras}, {Garcia-Gutierrez}, {Garofalo},
  {Jordi}, {Kervella}, {Khanna}, {Leclerc}, {Licata}, {Lambert}, {Marrese},
  {Masip}, {Ramos}, {Robichon}, {Robin}, {Romero-G{\'o}mez}, {Rubele}, \&
  {Weiler}}]{Fabricius_2021}
{Fabricius}, C., {Luri}, X., {Arenou}, F., {et~al.} 2021,
  \href{http://dx.doi.org/10.1051/0004-6361/202039834}{\color{magenta}\aap},
  \href{https://ui.adsabs.harvard.edu/abs/2021A&A...649A...5F}{\color{cyan}649},
  A5

\bibitem[{{Farihi} {et~al.}(2008){Farihi}, {Becklin}, \&
  {Zuckerman}}]{Farihi_2008}
{Farihi}, J., {Becklin}, E.~E., \& {Zuckerman}, B. 2008,
  \href{http://dx.doi.org/10.1086/588726}{\color{magenta}\apj},
  \href{https://ui.adsabs.harvard.edu/abs/2008ApJ...681.1470F}{\color{cyan}681},
  1470

\bibitem[{{Ferrario} {et~al.}(2015){Ferrario}, {de Martino}, \&
  {G{\"a}nsicke}}]{Ferrario_2015}
{Ferrario}, L., {de Martino}, D., \& {G{\"a}nsicke}, B.~T. 2015,
  \href{http://dx.doi.org/10.1007/s11214-015-0152-0}{\color{magenta}\ssr},
  \href{https://ui.adsabs.harvard.edu/abs/2015SSRv..191..111F}{\color{cyan}191},
  111

\bibitem[{{Fischer} \& {Valenti}(2005)}]{Fischer_2005}
{Fischer}, D.~A., \& {Valenti}, J. 2005,
  \href{http://dx.doi.org/10.1086/428383}{\color{magenta}\apj},
  \href{https://ui.adsabs.harvard.edu/abs/2005ApJ...622.1102F}{\color{cyan}622},
  1102

\bibitem[{{Fulton} {et~al.}(2021){Fulton}, {Rosenthal}, {Hirsch}, {Isaacson},
  {Howard}, {Dedrick}, {Sherstyuk}, {Blunt}, {Petigura}, {Knutson}, {Behmard},
  {Chontos}, {Crepp}, {Crossfield}, {Dalba}, {Fischer}, {Henry}, {Kane},
  {Kosiarek}, {Marcy}, {Rubenzahl}, {Weiss}, \& {Wright}}]{Fulton_2021}
{Fulton}, B.~J., {Rosenthal}, L.~J., {Hirsch}, L.~A., {et~al.} 2021,
  \href{http://dx.doi.org/10.3847/1538-4365/abfcc1}{\color{magenta}\apjs},
  \href{https://ui.adsabs.harvard.edu/abs/2021ApJS..255...14F}{\color{cyan}255},
  14

\bibitem[{{Gaia Collaboration} {et~al.}(2016){Gaia Collaboration}, {Prusti},
  {de Bruijne}, {Brown}, {Vallenari}, {Babusiaux}, {Bailer-Jones}, {Bastian},
  {Biermann}, {Evans}, {Eyer}, {Jansen}, {Jordi}, {Klioner}, {Lammers},
  {Lindegren}, {Luri}, {Mignard}, {Milligan}, {Panem}, {Poinsignon},
  {Pourbaix}, {Randich}, {Sarri}, {Sartoretti}, {Siddiqui}, {Soubiran},
  {Valette}, {van Leeuwen}, {Walton}, {Aerts}, {Arenou}, {Cropper}, {Drimmel},
  {H{\o}g}, {Katz}, {Lattanzi}, {O'Mullane}, {Grebel}, {Holland}, {Huc},
  {Passot}, {Bramante}, {Cacciari}, {Casta{\~n}eda}, {Chaoul}, {Cheek}, {De
  Angeli}, {Fabricius}, {Guerra}, {Hern{\'a}ndez}, {Jean-Antoine-Piccolo},
  {Masana}, {Messineo}, {Mowlavi}, {Nienartowicz}, {Ord{\'o}{\~n}ez-Blanco},
  {Panuzzo}, {Portell}, {Richards}, {Riello}, {Seabroke}, {Tanga},
  {Th{\'e}venin}, {Torra}, {Els}, {Gracia-Abril}, {Comoretto},
  {Garcia-Reinaldos}, {Lock}, {Mercier}, {Altmann}, {Andrae}, {Astraatmadja},
  {Bellas-Velidis}, {Benson}, {Berthier}, {Blomme}, {Busso}, {Carry},
  {Cellino}, {Clementini}, {Cowell}, {Creevey}, {Cuypers}, {Davidson}, {De
  Ridder}, {de Torres}, {Delchambre}, {Dell'Oro}, {Ducourant}, {Fr{\'e}mat},
  {Garc{\'\i}a-Torres}, {Gosset}, {Halbwachs}, {Hambly}, {Harrison}, {Hauser},
  {Hestroffer}, {Hodgkin}, {Huckle}, {Hutton}, {Jasniewicz}, {Jordan},
  {Kontizas}, {Korn}, {Lanzafame}, {Manteiga}, {Moitinho}, {Muinonen},
  {Osinde}, {Pancino}, {Pauwels}, {Petit}, {Recio-Blanco}, {Robin}, {Sarro},
  {Siopis}, {Smith}, {Smith}, {Sozzetti}, {Thuillot}, {van Reeven}, {Viala},
  {Abbas}, {Abreu Aramburu}, {Accart}, {Aguado}, {Allan}, {Allasia},
  {Altavilla}, {{\'A}lvarez}, {Alves}, {Anderson}, {Andrei}, {Anglada Varela},
  {Antiche}, {Antoja}, {Ant{\'o}n}, {Arcay}, {Atzei}, {Ayache}, {Bach},
  {Baker}, {Balaguer-N{\'u}{\~n}ez}, {Barache}, {Barata}, {Barbier}, {Barblan},
  {Baroni}, {Barrado y Navascu{\'e}s}, {Barros}, {Barstow}, {Becciani},
  {Bellazzini}, {Bellei}, {Bello Garc{\'\i}a}, {Belokurov}, {Bendjoya},
  {Berihuete}, {Bianchi}, {Bienaym{\'e}}, {Billebaud}, {Blagorodnova},
  {Blanco-Cuaresma}, {Boch}, {Bombrun}, {Borrachero}, {Bouquillon}, {Bourda},
  {Bouy}, {Bragaglia}, {Breddels}, {Brouillet}, {Br{\"u}semeister},
  {Bucciarelli}, {Budnik}, {Burgess}, {Burgon}, {Burlacu}, {Busonero}, {Buzzi},
  {Caffau}, {Cambras}, {Campbell}, {Cancelliere}, {Cantat-Gaudin}, {Carlucci},
  {Carrasco}, {Castellani}, {Charlot}, {Charnas}, {Charvet}, {Chassat},
  {Chiavassa}, {Clotet}, {Cocozza}, {Collins}, {Collins}, {Costigan}, {Crifo},
  {Cross}, {Crosta}, {Crowley}, {Dafonte}, {Damerdji}, {Dapergolas}, {David},
  {David}, {De Cat}, {de Felice}, {de Laverny}, {De Luise}, {De March}, {de
  Martino}, {de Souza}, {Debosscher}, {del Pozo}, {Delbo}, {Delgado},
  {Delgado}, {di Marco}, {Di Matteo}, {Diakite}, {Distefano}, {Dolding}, {Dos
  Anjos}, {Drazinos}, {Dur{\'a}n}, {Dzigan}, {Ecale}, {Edvardsson}, {Enke},
  {Erdmann}, {Escolar}, {Espina}, {Evans}, {Eynard Bontemps}, {Fabre},
  {Fabrizio}, {Faigler}, {Falc{\~a}o}, {Farr{\`a}s Casas}, {Faye}, {Federici},
  {Fedorets}, {Fern{\'a}ndez-Hern{\'a}ndez}, {Fernique}, {Fienga}, {Figueras},
  {Filippi}, {Findeisen}, {Fonti}, {Fouesneau}, {Fraile}, {Fraser}, {Fuchs},
  {Furnell}, {Gai}, {Galleti}, {Galluccio}, {Garabato}, {Garc{\'\i}a-Sedano},
  {Gar{\'e}}, {Garofalo}, {Garralda}, {Gavras}, {Gerssen}, {Geyer}, {Gilmore},
  {Girona}, {Giuffrida}, {Gomes}, {Gonz{\'a}lez-Marcos},
  {Gonz{\'a}lez-N{\'u}{\~n}ez}, {Gonz{\'a}lez-Vidal}, {Granvik}, {Guerrier},
  {Guillout}, {Guiraud}, {G{\'u}rpide}, {Guti{\'e}rrez-S{\'a}nchez}, {Guy},
  {Haigron}, {Hatzidimitriou}, {Haywood}, {Heiter}, {Helmi}, {Hobbs},
  {Hofmann}, {Holl}, {Holland}, {Hunt}, {Hypki}, {Icardi}, {Irwin}, {Jevardat
  de Fombelle}, {Jofr{\'e}}, {Jonker}, {Jorissen}, {Julbe}, {Karampelas},
  {Kochoska}, {Kohley}, {Kolenberg}, {Kontizas}, {Koposov}, {Kordopatis},
  {Koubsky}, {Kowalczyk}, {Krone-Martins}, {Kudryashova}, {Kull}, {Bachchan},
  {Lacoste-Seris}, {Lanza}, {Lavigne}, {Le Poncin-Lafitte}, {Lebreton},
  {Lebzelter}, {Leccia}, {Leclerc}, {Lecoeur-Taibi}, {Lemaitre}, {Lenhardt},
  {Leroux}, {Liao}, {Licata}, {Lindstr{\o}m}, {Lister}, {Livanou}, {Lobel},
  {L{\"o}ffler}, {L{\'o}pez}, {Lopez-Lozano}, {Lorenz}, {Loureiro},
  {MacDonald}, {Magalh{\~a}es Fernandes}, {Managau}, {Mann}, {Mantelet},
  {Marchal}, {Marchant}, {Marconi}, {Marie}, {Marinoni}, {Marrese},
  {Marschalk{\'o}}, {Marshall}, {Mart{\'\i}n-Fleitas}, {Martino}, {Mary},
  {Matijevi{\v{c}}}, {Mazeh}, {McMillan}, {Messina}, {Mestre}, {Michalik},
  {Millar}, {Miranda}, {Molina}, {Molinaro}, {Molinaro}, {Moln{\'a}r},
  {Moniez}, {Montegriffo}, {Monteiro}, {Mor}, {Mora}, {Morbidelli}, {Morel},
  {Morgenthaler}, {Morley}, {Morris}, {Mulone}, {Muraveva}, {Musella},
  {Narbonne}, {Nelemans}, {Nicastro}, {Noval}, {Ord{\'e}novic},
  {Ordieres-Mer{\'e}}, {Osborne}, {Pagani}, {Pagano}, {Pailler}, {Palacin},
  {Palaversa}, {Parsons}, {Paulsen}, {Pecoraro}, {Pedrosa}, {Pentik{\"a}inen},
  {Pereira}, {Pichon}, {Piersimoni}, {Pineau}, {Plachy}, {Plum}, {Poujoulet},
  {Pr{\v{s}}a}, {Pulone}, {Ragaini}, {Rago}, {Rambaux}, {Ramos-Lerate},
  {Ranalli}, {Rauw}, {Read}, {Regibo}, {Renk}, {Reyl{\'e}}, {Ribeiro},
  {Rimoldini}, {Ripepi}, {Riva}, {Rixon}, {Roelens}, {Romero-G{\'o}mez},
  {Rowell}, {Royer}, {Rudolph}, {Ruiz-Dern}, {Sadowski}, {Sagrist{\`a}
  Sell{\'e}s}, {Sahlmann}, {Salgado}, {Salguero}, {Sarasso}, {Savietto},
  {Schnorhk}, {Schultheis}, {Sciacca}, {Segol}, {Segovia}, {Segransan},
  {Serpell}, {Shih}, {Smareglia}, {Smart}, {Smith}, {Solano}, {Solitro},
  {Sordo}, {Soria Nieto}, {Souchay}, {Spagna}, {Spoto}, {Stampa}, {Steele},
  {Steidelm{\"u}ller}, {Stephenson}, {Stoev}, {Suess}, {S{\"u}veges}, {Surdej},
  {Szabados}, {Szegedi-Elek}, {Tapiador}, {Taris}, {Tauran}, {Taylor},
  {Teixeira}, {Terrett}, {Tingley}, {Trager}, {Turon}, {Ulla}, {Utrilla},
  {Valentini}, {van Elteren}, {Van Hemelryck}, {van Leeuwen}, {Varadi},
  {Vecchiato}, {Veljanoski}, {Via}, {Vicente}, {Vogt}, {Voss}, {Votruba},
  {Voutsinas}, {Walmsley}, {Weiler}, {Weingrill}, {Werner}, {Wevers},
  {Whitehead}, {Wyrzykowski}, {Yoldas}, {{\v{Z}}erjal}, {Zucker}, {Zurbach},
  {Zwitter}, {Alecu}, {Allen}, {Allende Prieto}, {Amorim},
  {Anglada-Escud{\'e}}, {Arsenijevic}, {Azaz}, {Balm}, {Beck}, {Bernstein},
  {Bigot}, {Bijaoui}, {Blasco}, {Bonfigli}, {Bono}, {Boudreault}, {Bressan},
  {Brown}, {Brunet}, {Bunclark}, {Buonanno}, {Butkevich}, {Carret}, {Carrion},
  {Chemin}, {Ch{\'e}reau}, {Corcione}, {Darmigny}, {de Boer}, {de Teodoro}, {de
  Zeeuw}, {Delle Luche}, {Domingues}, {Dubath}, {Fodor}, {Fr{\'e}zouls},
  {Fries}, {Fustes}, {Fyfe}, {Gallardo}, {Gallegos}, {Gardiol}, {Gebran},
  {Gomboc}, {G{\'o}mez}, {Grux}, {Gueguen}, {Heyrovsky}, {Hoar}, {Iannicola},
  {Isasi Parache}, {Janotto}, {Joliet}, {Jonckheere}, {Keil}, {Kim},
  {Klagyivik}, {Klar}, {Knude}, {Kochukhov}, {Kolka}, {Kos}, {Kutka}, {Lainey},
  {LeBouquin}, {Liu}, {Loreggia}, {Makarov}, {Marseille}, {Martayan},
  {Martinez-Rubi}, {Massart}, {Meynadier}, {Mignot}, {Munari}, {Nguyen},
  {Nordlander}, {Ocvirk}, {O'Flaherty}, {Olias Sanz}, {Ortiz}, {Osorio},
  {Oszkiewicz}, {Ouzounis}, {Palmer}, {Park}, {Pasquato}, {Peltzer}, {Peralta},
  {P{\'e}turaud}, {Pieniluoma}, {Pigozzi}, {Poels}, {Prat}, {Prod'homme},
  {Raison}, {Rebordao}, {Risquez}, {Rocca-Volmerange}, {Rosen}, {Ruiz-Fuertes},
  {Russo}, {Sembay}, {Serraller Vizcaino}, {Short}, {Siebert}, {Silva},
  {Sinachopoulos}, {Slezak}, {Soffel}, {Sosnowska}, {Strai{\v{z}}ys}, {ter
  Linden}, {Terrell}, {Theil}, {Tiede}, {Troisi}, {Tsalmantza}, {Tur},
  {Vaccari}, {Vachier}, {Valles}, {Van Hamme}, {Veltz}, {Virtanen}, {Wallut},
  {Wichmann}, {Wilkinson}, {Ziaeepour}, \& {Zschocke}}]{GaiaCollaboration_2016}
{Gaia Collaboration}, {Prusti}, T., {de Bruijne}, J.~H.~J., {et~al.} 2016,
  \href{http://dx.doi.org/10.1051/0004-6361/201629272}{\color{magenta}\aap},
  \href{https://ui.adsabs.harvard.edu/abs/2016A&A...595A...1G}{\color{cyan}595},
  A1

\bibitem[{{Gaia Collaboration} {et~al.}(2021{\natexlab{a}}){Gaia
  Collaboration}, {Smart}, {Sarro}, {Rybizki}, {Reyl{\'e}}, {Robin}, {Hambly},
  {Abbas}, {Barstow}, {de Bruijne}, {Bucciarelli}, {Carrasco}, {Cooper},
  {Hodgkin}, {Masana}, {Michalik}, {Sahlmann}, {Sozzetti}, {Brown},
  {Vallenari}, {Prusti}, {Babusiaux}, {Biermann}, {Creevey}, {Evans}, {Eyer},
  {Hutton}, {Jansen}, {Jordi}, {Klioner}, {Lammers}, {Lindegren}, {Luri},
  {Mignard}, {Panem}, {Pourbaix}, {Randich}, {Sartoretti}, {Soubiran},
  {Walton}, {Arenou}, {Bailer-Jones}, {Bastian}, {Cropper}, {Drimmel}, {Katz},
  {Lattanzi}, {van Leeuwen}, {Bakker}, {Casta{\~n}eda}, {De Angeli},
  {Ducourant}, {Fabricius}, {Fouesneau}, {Fr{\'e}mat}, {Guerra}, {Guerrier},
  {Guiraud}, {Jean-Antoine Piccolo}, {Messineo}, {Mowlavi}, {Nicolas},
  {Nienartowicz}, {Pailler}, {Panuzzo}, {Riclet}, {Roux}, {Seabroke}, {Sordo},
  {Tanga}, {Th{\'e}venin}, {Gracia-Abril}, {Portell}, {Teyssier}, {Altmann},
  {Andrae}, {Bellas-Velidis}, {Benson}, {Berthier}, {Blomme}, {Brugaletta},
  {Burgess}, {Busso}, {Carry}, {Cellino}, {Cheek}, {Clementini}, {Damerdji},
  {Davidson}, {Delchambre}, {Dell'Oro}, {Fern{\'a}ndez-Hern{\'a}ndez},
  {Galluccio}, {Garc{\'\i}a-Lario}, {Garcia-Reinaldos},
  {Gonz{\'a}lez-N{\'u}{\~n}ez}, {Gosset}, {Haigron}, {Halbwachs}, {Harrison},
  {Hatzidimitriou}, {Heiter}, {Hern{\'a}ndez}, {Hestroffer}, {Holl},
  {Jan{\ss}en}, {Jevardat de Fombelle}, {Jordan}, {Krone-Martins}, {Lanzafame},
  {L{\"o}ffler}, {Lorca}, {Manteiga}, {Marchal}, {Marrese}, {Moitinho}, {Mora},
  {Muinonen}, {Osborne}, {Pancino}, {Pauwels}, {Recio-Blanco}, {Richards},
  {Riello}, {Rimoldini}, {Roegiers}, {Siopis}, {Smith}, {Ulla}, {Utrilla}, {van
  Leeuwen}, {van Reeven}, {Abreu Aramburu}, {Accart}, {Aerts}, {Aguado},
  {Ajaj}, {Altavilla}, {{\'A}lvarez}, {{\'A}lvarez Cid-Fuentes}, {Alves},
  {Anderson}, {Anglada Varela}, {Antoja}, {Audard}, {Baines}, {Baker},
  {Balaguer-N{\'u}{\~n}ez}, {Balbinot}, {Balog}, {Barache}, {Barbato},
  {Barros}, {Bartolom{\'e}}, {Bassilana}, {Bauchet}, {Baudesson-Stella},
  {Becciani}, {Bellazzini}, {Bernet}, {Bertone}, {Bianchi}, {Blanco-Cuaresma},
  {Boch}, {Bombrun}, {Bossini}, {Bouquillon}, {Bragaglia}, {Bramante},
  {Breedt}, {Bressan}, {Brouillet}, {Burlacu}, {Busonero}, {Butkevich},
  {Buzzi}, {Caffau}, {Cancelliere}, {C{\'a}novas}, {Cantat-Gaudin}, {Carballo},
  {Carlucci}, {Carnerero}, {Casamiquela}, {Castellani}, {Castro-Ginard},
  {Castro Sampol}, {Chaoul}, {Charlot}, {Chemin}, {Chiavassa}, {Cioni},
  {Comoretto}, {Cornez}, {Cowell}, {Crifo}, {Crosta}, {Crowley}, {Dafonte},
  {Dapergolas}, {David}, {David}, {de Laverny}, {De Luise}, {De March}, {De
  Ridder}, {de Souza}, {de Teodoro}, {de Torres}, {del Peloso}, {del Pozo},
  {Delgado}, {Delgado}, {Delisle}, {Di Matteo}, {Diakite}, {Diener},
  {Distefano}, {Dolding}, {Eappachen}, {Edvardsson}, {Enke}, {Esquej}, {Fabre},
  {Fabrizio}, {Faigler}, {Fedorets}, {Fernique}, {Fienga}, {Figueras},
  {Fouron}, {Fragkoudi}, {Fraile}, {Franke}, {Gai}, {Garabato},
  {Garcia-Gutierrez}, {Garc{\'\i}a-Torres}, {Garofalo}, {Gavras}, {Gerlach},
  {Geyer}, {Giacobbe}, {Gilmore}, {Girona}, {Giuffrida}, {Gomel}, {Gomez},
  {Gonzalez-Santamaria}, {Gonz{\'a}lez-Vidal}, {Granvik},
  {Guti{\'e}rrez-S{\'a}nchez}, {Guy}, {Hauser}, {Haywood}, {Helmi}, {Hidalgo},
  {Hilger}, {H{\l}adczuk}, {Hobbs}, {Holland}, {Huckle}, {Jasniewicz},
  {Jonker}, {Juaristi Campillo}, {Julbe}, {Karbevska}, {Kervella}, {Khanna},
  {Kochoska}, {Kontizas}, {Kordopatis}, {Korn}, {Kostrzewa-Rutkowska},
  {Kruszy{\'n}ska}, {Lambert}, {Lanza}, {Lasne}, {Le Campion}, {Le Fustec},
  {Lebreton}, {Lebzelter}, {Leccia}, {Leclerc}, {Lecoeur-Taibi}, {Liao},
  {Licata}, {Lindstr{\o}m}, {Lister}, {Livanou}, {Lobel}, {Madrero Pardo},
  {Managau}, {Mann}, {Marchant}, {Marconi}, {Marcos Santos}, {Marinoni},
  {Marocco}, {Marshall}, {Martin Polo}, {Mart{\'\i}n-Fleitas}, {Masip},
  {Massari}, {Mastrobuono-Battisti}, {Mazeh}, {McMillan}, {Messina}, {Millar},
  {Mints}, {Molina}, {Molinaro}, {Moln{\'a}r}, {Montegriffo}, {Mor},
  {Morbidelli}, {Morel}, {Morris}, {Mulone}, {Munoz}, {Muraveva}, {Murphy},
  {Musella}, {Noval}, {Ord{\'e}novic}, {Orr{\`u}}, {Osinde}, {Pagani},
  {Pagano}, {Palaversa}, {Palicio}, {Panahi}, {Pawlak}, {Pe{\~n}alosa
  Esteller}, {Penttil{\"a}}, {Piersimoni}, {Pineau}, {Plachy}, {Plum},
  {Poggio}, {Poretti}, {Poujoulet}, {Pr{\v{s}}a}, {Pulone}, {Racero},
  {Ragaini}, {Rainer}, {Raiteri}, {Rambaux}, {Ramos}, {Ramos-Lerate}, {Re
  Fiorentin}, {Regibo}, {Ripepi}, {Riva}, {Rixon}, {Robichon}, {Robin},
  {Roelens}, {Rohrbasser}, {Romero-G{\'o}mez}, {Rowell}, {Royer}, {Rybicki},
  {Sadowski}, {Sagrist{\`a} Sell{\'e}s}, {Salgado}, {Salguero}, {Samaras},
  {Sanchez Gimenez}, {Sanna}, {Santove{\~n}a}, {Sarasso}, {Schultheis},
  {Sciacca}, {Segol}, {Segovia}, {S{\'e}gransan}, {Semeux}, {Shahaf},
  {Siddiqui}, {Siebert}, {Siltala}, {Slezak}, {Solano}, {Solitro}, {Souami},
  {Souchay}, {Spagna}, {Spoto}, {Steele}, {Steidelm{\"u}ller}, {Stephenson},
  {S{\"u}veges}, {Szabados}, {Szegedi-Elek}, {Taris}, {Tauran}, {Taylor},
  {Teixeira}, {Thuillot}, {Tonello}, {Torra}, {Torra}, {Turon}, {Unger},
  {Vaillant}, {van Dillen}, {Vanel}, {Vecchiato}, {Viala}, {Vicente},
  {Voutsinas}, {Weiler}, {Wevers}, {Wyrzykowski}, {Yoldas}, {Yvard}, {Zhao},
  {Zorec}, {Zucker}, {Zurbach}, \& {Zwitter}}]{GaiaCollaboration_2021b}
{Gaia Collaboration}, {Smart}, R.~L., {Sarro}, L.~M., {et~al.}
  2021{\natexlab{a}},
  \href{http://dx.doi.org/10.1051/0004-6361/202039498}{\color{magenta}\aap},
  \href{https://ui.adsabs.harvard.edu/abs/2021A&A...649A...6G}{\color{cyan}649},
  A6

\bibitem[{{Gaia Collaboration} {et~al.}(2021{\natexlab{b}}){Gaia
  Collaboration}, {Brown}, {Vallenari}, {Prusti}, {de Bruijne}, {Babusiaux},
  {Biermann}, {Creevey}, {Evans}, {Eyer}, {Hutton}, {Jansen}, {Jordi},
  {Klioner}, {Lammers}, {Lindegren}, {Luri}, {Mignard}, {Panem}, {Pourbaix},
  {Randich}, {Sartoretti}, {Soubiran}, {Walton}, {Arenou}, {Bailer-Jones},
  {Bastian}, {Cropper}, {Drimmel}, {Katz}, {Lattanzi}, {van Leeuwen}, {Bakker},
  {Cacciari}, {Casta{\~n}eda}, {De Angeli}, {Ducourant}, {Fabricius},
  {Fouesneau}, {Fr{\'e}mat}, {Guerra}, {Guerrier}, {Guiraud}, {Jean-Antoine
  Piccolo}, {Masana}, {Messineo}, {Mowlavi}, {Nicolas}, {Nienartowicz},
  {Pailler}, {Panuzzo}, {Riclet}, {Roux}, {Seabroke}, {Sordo}, {Tanga},
  {Th{\'e}venin}, {Gracia-Abril}, {Portell}, {Teyssier}, {Altmann}, {Andrae},
  {Bellas-Velidis}, {Benson}, {Berthier}, {Blomme}, {Brugaletta}, {Burgess},
  {Busso}, {Carry}, {Cellino}, {Cheek}, {Clementini}, {Damerdji}, {Davidson},
  {Delchambre}, {Dell'Oro}, {Fern{\'a}ndez-Hern{\'a}ndez}, {Galluccio},
  {Garc{\'\i}a-Lario}, {Garcia-Reinaldos}, {Gonz{\'a}lez-N{\'u}{\~n}ez},
  {Gosset}, {Haigron}, {Halbwachs}, {Hambly}, {Harrison}, {Hatzidimitriou},
  {Heiter}, {Hern{\'a}ndez}, {Hestroffer}, {Hodgkin}, {Holl}, {Jan{\ss}en},
  {Jevardat de Fombelle}, {Jordan}, {Krone-Martins}, {Lanzafame},
  {L{\"o}ffler}, {Lorca}, {Manteiga}, {Marchal}, {Marrese}, {Moitinho}, {Mora},
  {Muinonen}, {Osborne}, {Pancino}, {Pauwels}, {Petit}, {Recio-Blanco},
  {Richards}, {Riello}, {Rimoldini}, {Robin}, {Roegiers}, {Rybizki}, {Sarro},
  {Siopis}, {Smith}, {Sozzetti}, {Ulla}, {Utrilla}, {van Leeuwen}, {van
  Reeven}, {Abbas}, {Abreu Aramburu}, {Accart}, {Aerts}, {Aguado}, {Ajaj},
  {Altavilla}, {{\'A}lvarez}, {{\'A}lvarez Cid-Fuentes}, {Alves}, {Anderson},
  {Anglada Varela}, {Antoja}, {Audard}, {Baines}, {Baker},
  {Balaguer-N{\'u}{\~n}ez}, {Balbinot}, {Balog}, {Barache}, {Barbato},
  {Barros}, {Barstow}, {Bartolom{\'e}}, {Bassilana}, {Bauchet},
  {Baudesson-Stella}, {Becciani}, {Bellazzini}, {Bernet}, {Bertone}, {Bianchi},
  {Blanco-Cuaresma}, {Boch}, {Bombrun}, {Bossini}, {Bouquillon}, {Bragaglia},
  {Bramante}, {Breedt}, {Bressan}, {Brouillet}, {Bucciarelli}, {Burlacu},
  {Busonero}, {Butkevich}, {Buzzi}, {Caffau}, {Cancelliere}, {C{\'a}novas},
  {Cantat-Gaudin}, {Carballo}, {Carlucci}, {Carnerero}, {Carrasco},
  {Casamiquela}, {Castellani}, {Castro-Ginard}, {Castro Sampol}, {Chaoul},
  {Charlot}, {Chemin}, {Chiavassa}, {Cioni}, {Comoretto}, {Cooper}, {Cornez},
  {Cowell}, {Crifo}, {Crosta}, {Crowley}, {Dafonte}, {Dapergolas}, {David},
  {David}, {de Laverny}, {De Luise}, {De March}, {De Ridder}, {de Souza}, {de
  Teodoro}, {de Torres}, {del Peloso}, {del Pozo}, {Delbo}, {Delgado},
  {Delgado}, {Delisle}, {Di Matteo}, {Diakite}, {Diener}, {Distefano},
  {Dolding}, {Eappachen}, {Edvardsson}, {Enke}, {Esquej}, {Fabre}, {Fabrizio},
  {Faigler}, {Fedorets}, {Fernique}, {Fienga}, {Figueras}, {Fouron},
  {Fragkoudi}, {Fraile}, {Franke}, {Gai}, {Garabato}, {Garcia-Gutierrez},
  {Garc{\'\i}a-Torres}, {Garofalo}, {Gavras}, {Gerlach}, {Geyer}, {Giacobbe},
  {Gilmore}, {Girona}, {Giuffrida}, {Gomel}, {Gomez}, {Gonzalez-Santamaria},
  {Gonz{\'a}lez-Vidal}, {Granvik}, {Guti{\'e}rrez-S{\'a}nchez}, {Guy},
  {Hauser}, {Haywood}, {Helmi}, {Hidalgo}, {Hilger}, {H{\l}adczuk}, {Hobbs},
  {Holland}, {Huckle}, {Jasniewicz}, {Jonker}, {Juaristi Campillo}, {Julbe},
  {Karbevska}, {Kervella}, {Khanna}, {Kochoska}, {Kontizas}, {Kordopatis},
  {Korn}, {Kostrzewa-Rutkowska}, {Kruszy{\'n}ska}, {Lambert}, {Lanza}, {Lasne},
  {Le Campion}, {Le Fustec}, {Lebreton}, {Lebzelter}, {Leccia}, {Leclerc},
  {Lecoeur-Taibi}, {Liao}, {Licata}, {Lindstr{\o}m}, {Lister}, {Livanou},
  {Lobel}, {Madrero Pardo}, {Managau}, {Mann}, {Marchant}, {Marconi}, {Marcos
  Santos}, {Marinoni}, {Marocco}, {Marshall}, {Martin Polo},
  {Mart{\'\i}n-Fleitas}, {Masip}, {Massari}, {Mastrobuono-Battisti}, {Mazeh},
  {McMillan}, {Messina}, {Michalik}, {Millar}, {Mints}, {Molina}, {Molinaro},
  {Moln{\'a}r}, {Montegriffo}, {Mor}, {Morbidelli}, {Morel}, {Morris},
  {Mulone}, {Munoz}, {Muraveva}, {Murphy}, {Musella}, {Noval}, {Ord{\'e}novic},
  {Orr{\`u}}, {Osinde}, {Pagani}, {Pagano}, {Palaversa}, {Palicio}, {Panahi},
  {Pawlak}, {Pe{\~n}alosa Esteller}, {Penttil{\"a}}, {Piersimoni}, {Pineau},
  {Plachy}, {Plum}, {Poggio}, {Poretti}, {Poujoulet}, {Pr{\v{s}}a}, {Pulone},
  {Racero}, {Ragaini}, {Rainer}, {Raiteri}, {Rambaux}, {Ramos}, {Ramos-Lerate},
  {Re Fiorentin}, {Regibo}, {Reyl{\'e}}, {Ripepi}, {Riva}, {Rixon}, {Robichon},
  {Robin}, {Roelens}, {Rohrbasser}, {Romero-G{\'o}mez}, {Rowell}, {Royer},
  {Rybicki}, {Sadowski}, {Sagrist{\`a} Sell{\'e}s}, {Sahlmann}, {Salgado},
  {Salguero}, {Samaras}, {Sanchez Gimenez}, {Sanna}, {Santove{\~n}a},
  {Sarasso}, {Schultheis}, {Sciacca}, {Segol}, {Segovia}, {S{\'e}gransan},
  {Semeux}, {Shahaf}, {Siddiqui}, {Siebert}, {Siltala}, {Slezak}, {Smart},
  {Solano}, {Solitro}, {Souami}, {Souchay}, {Spagna}, {Spoto}, {Steele},
  {Steidelm{\"u}ller}, {Stephenson}, {S{\"u}veges}, {Szabados}, {Szegedi-Elek},
  {Taris}, {Tauran}, {Taylor}, {Teixeira}, {Thuillot}, {Tonello}, {Torra},
  {Torra}, {Turon}, {Unger}, {Vaillant}, {van Dillen}, {Vanel}, {Vecchiato},
  {Viala}, {Vicente}, {Voutsinas}, {Weiler}, {Wevers}, {Wyrzykowski}, {Yoldas},
  {Yvard}, {Zhao}, {Zorec}, {Zucker}, {Zurbach}, \&
  {Zwitter}}]{GaiaCollaboration_2021a}
{Gaia Collaboration}, {Brown}, A.~G.~A., {Vallenari}, A., {et~al.}
  2021{\natexlab{b}},
  \href{http://dx.doi.org/10.1051/0004-6361/202039657}{\color{magenta}\aap},
  \href{https://ui.adsabs.harvard.edu/abs/2021A&A...649A...1G}{\color{cyan}649},
  A1

\bibitem[{{Gentile Fusillo} {et~al.}(2019){Gentile Fusillo}, {Tremblay},
  {G{\"a}nsicke}, {Manser}, {Cunningham}, {Cukanovaite}, {Hollands}, {Marsh},
  {Raddi}, {Jordan}, {Toonen}, {Geier}, {Barstow}, \&
  {Cummings}}]{Gentile_Fusillo_2019}
{Gentile Fusillo}, N.~P., {Tremblay}, P.-E., {G{\"a}nsicke}, B.~T., {et~al.}
  2019, \href{http://dx.doi.org/10.1093/mnras/sty3016}{\color{magenta}\mnras},
  \href{https://ui.adsabs.harvard.edu/abs/2019MNRAS.482.4570G}{\color{cyan}482},
  4570

\bibitem[{{Ghezzi} {et~al.}(2018){Ghezzi}, {Montet}, \&
  {Johnson}}]{Ghezzi_2018}
{Ghezzi}, L., {Montet}, B.~T., \& {Johnson}, J.~A. 2018,
  \href{http://dx.doi.org/10.3847/1538-4357/aac37c}{\color{magenta}\apj},
  \href{https://ui.adsabs.harvard.edu/abs/2018ApJ...860..109G}{\color{cyan}860},
  109

\bibitem[{{Gould} \& {Kilic}(2008)}]{Gould_2008}
{Gould}, A., \& {Kilic}, M. 2008,
  \href{http://dx.doi.org/10.1086/527476}{\color{magenta}\apjl},
  \href{https://ui.adsabs.harvard.edu/abs/2008ApJ...673L..75G}{\color{cyan}673},
  L75

\bibitem[{{Harding} {et~al.}(2016){Harding}, {Hallinan}, {Milburn}, {Gardner},
  {Konidaris}, {Singh}, {Shao}, {Sandhu}, {Kyne}, \&
  {Schlichting}}]{Harding_2016}
{Harding}, L.~K., {Hallinan}, G., {Milburn}, J., {et~al.} 2016,
  \href{http://dx.doi.org/10.1093/mnras/stw094}{\color{magenta}\mnras},
  \href{https://ui.adsabs.harvard.edu/abs/2016MNRAS.457.3036H}{\color{cyan}457},
  3036

\bibitem[{Harris {et~al.}(2020)Harris, Millman, van~der Walt, Gommers,
  Virtanen, Cournapeau, Wieser, Taylor, Berg, Smith, Kern, Picus, Hoyer, van
  Kerkwijk, Brett, Haldane, del R{\'{i}}o, Wiebe, Peterson,
  G{\'{e}}rard-Marchant, Sheppard, Reddy, Weckesser, Abbasi, Gohlke, \&
  Oliphant}]{harris2020array}
Harris, C.~R., Millman, K.~J., van~der Walt, S.~J., {et~al.} 2020,
  \href{http://dx.doi.org/10.1038/s41586-020-2649-2}{\color{magenta}Nature},
  585, 357

\bibitem[{{Hekker} \& {Mel{\'e}ndez}(2007)}]{Hekker_2007}
{Hekker}, S., \& {Mel{\'e}ndez}, J. 2007,
  \href{http://dx.doi.org/10.1051/0004-6361:20078233}{\color{magenta}\aap},
  \href{https://ui.adsabs.harvard.edu/abs/2007A&A...475.1003H}{\color{cyan}475},
  1003

\bibitem[{{Hogan} {et~al.}(2009){Hogan}, {Burleigh}, \& {Clarke}}]{Hogan_2009}
{Hogan}, E., {Burleigh}, M.~R., \& {Clarke}, F.~J. 2009,
  \href{http://dx.doi.org/10.1111/j.1365-2966.2009.14565.x}{\color{magenta}\mnras},
  \href{https://ui.adsabs.harvard.edu/abs/2009MNRAS.396.2074H}{\color{cyan}396},
  2074

\bibitem[{{Hogan} {et~al.}(2010){Hogan}, {Burleigh}, \& {Clarke}}]{Hogan_2010}
{Hogan}, E., {Burleigh}, M.~R., \& {Clarke}, F.~J. 2010, in
  \href{http://dx.doi.org/10.1063/1.3527859}{\color{magenta}American Institute
  of Physics Conference Series}, Vol.
  \href{https://ui.adsabs.harvard.edu/abs/2010AIPC.1273..440H}{\color{cyan}1273},
  17th European White Dwarf Workshop, ed. K.~{Werner} \& T.~{Rauch}, 440

\bibitem[{{Hogan} {et~al.}(2011){Hogan}, {Burleigh}, \& {Clarke}}]{Hogan_2011}
{Hogan}, E., {Burleigh}, M.~R., \& {Clarke}, F.~J. 2011, in
  \href{http://dx.doi.org/10.1063/1.3556210}{\color{magenta}American Institute
  of Physics Conference Series}, Vol.
  \href{https://ui.adsabs.harvard.edu/abs/2011AIPC.1331..271H}{\color{cyan}1331},
  Planetary Systems Beyond the Main Sequence, ed. S.~{Schuh}, H.~{Drechsel}, \&
  U.~{Heber}, 271

\bibitem[{{Holberg} \& {Bergeron}(2006)}]{Holberg_2006}
{Holberg}, J.~B., \& {Bergeron}, P. 2006,
  \href{http://dx.doi.org/10.1086/505938}{\color{magenta}\aj},
  \href{https://ui.adsabs.harvard.edu/abs/2006AJ....132.1221H}{\color{cyan}132},
  1221

\bibitem[{{Holmberg} {et~al.}(2009){Holmberg}, {Nordstr{\"o}m}, \&
  {Andersen}}]{Holmberg_2009}
{Holmberg}, J., {Nordstr{\"o}m}, B., \& {Andersen}, J. 2009,
  \href{http://dx.doi.org/10.1051/0004-6361/200811191}{\color{magenta}\aap},
  \href{https://ui.adsabs.harvard.edu/abs/2009A&A...501..941H}{\color{cyan}501},
  941

\bibitem[{{Hubickyj} {et~al.}(2005){Hubickyj}, {Bodenheimer}, \&
  {Lissauer}}]{Hubickyj_2005}
{Hubickyj}, O., {Bodenheimer}, P., \& {Lissauer}, J.~J. 2005,
  \href{http://dx.doi.org/10.1016/j.icarus.2005.06.021}{\color{magenta}\icarus},
  \href{https://ui.adsabs.harvard.edu/abs/2005Icar..179..415H}{\color{cyan}179},
  415

\bibitem[{Hunter(2007)}]{Hunter_2007}
Hunter, J.~D. 2007,
  \href{http://dx.doi.org/10.1109/MCSE.2007.55}{\color{magenta}Computing in
  Science \& Engineering}, 9, 90

\bibitem[{{Ida} \& {Lin}(2005)}]{Ida_2005}
{Ida}, S., \& {Lin}, D.~N.~C. 2005,
  \href{http://dx.doi.org/10.1086/429953}{\color{magenta}\apj},
  \href{https://ui.adsabs.harvard.edu/abs/2005ApJ...626.1045I}{\color{cyan}626},
  1045

\bibitem[{{Janson} {et~al.}(2019){Janson}, {Asensio-Torres}, {Andr{\'e}},
  {Bonnefoy}, {Delorme}, {Reffert}, {Desidera}, {Langlois}, {Chauvin},
  {Gratton}, {Bohn}, {Eriksson}, {Marleau}, {Mamajek}, {Vigan}, \&
  {Carson}}]{Janson_2019}
{Janson}, M., {Asensio-Torres}, R., {Andr{\'e}}, D., {et~al.} 2019,
  \href{http://dx.doi.org/10.1051/0004-6361/201935687}{\color{magenta}\aap},
  \href{https://ui.adsabs.harvard.edu/abs/2019A&A...626A..99J}{\color{cyan}626},
  A99

\bibitem[{{Janson} {et~al.}(2021{\natexlab{a}}){Janson}, {Squicciarini},
  {Delorme}, {Gratton}, {Bonnefoy}, {Reffert}, {Mamajek}, {Eriksson}, {Vigan},
  {Langlois}, {Engler}, {Chauvin}, {Desidera}, {Mayer}, {Marleau}, {Bohn},
  {Samland}, {Meyer}, {d'Orazi}, {Henning}, {Quanz}, {Kenworthy}, \&
  {Carson}}]{Janson_2021}
{Janson}, M., {Squicciarini}, V., {Delorme}, P., {et~al.} 2021{\natexlab{a}},
  \href{http://dx.doi.org/10.1051/0004-6361/202039683}{\color{magenta}\aap},
  \href{https://ui.adsabs.harvard.edu/abs/2021A&A...646A.164J}{\color{cyan}646},
  A164

\bibitem[{{Janson} {et~al.}(2021{\natexlab{b}}){Janson}, {Gratton}, {Rodet},
  {Vigan}, {Bonnefoy}, {Delorme}, {Mamajek}, {Reffert}, {Stock}, {Marleau},
  {Langlois}, {Chauvin}, {Desidera}, {Ringqvist}, {Mayer}, {Viswanath},
  {Squicciarini}, {Meyer}, {Samland}, {Petrus}, {Helled}, {Kenworthy}, {Quanz},
  {Biller}, {Henning}, {Mesa}, {Engler}, \& {Carson}}]{Janson_2021nature}
{Janson}, M., {Gratton}, R., {Rodet}, L., {et~al.} 2021{\natexlab{b}},
  \href{http://dx.doi.org/10.1038/s41586-021-04124-8}{\color{magenta}\nat},
  \href{https://ui.adsabs.harvard.edu/abs/2021Natur.600..231J}{\color{cyan}600},
  231

\bibitem[{{Johnson} {et~al.}(2010){Johnson}, {Aller}, {Howard}, \&
  {Crepp}}]{Johnson_2010}
{Johnson}, J.~A., {Aller}, K.~M., {Howard}, A.~W., \& {Crepp}, J.~R. 2010,
  \href{http://dx.doi.org/10.1086/655775}{\color{magenta}\pasp},
  \href{https://ui.adsabs.harvard.edu/abs/2010PASP..122..905J}{\color{cyan}122},
  905

\bibitem[{{Jones} {et~al.}(2016){Jones}, {Jenkins}, {Brahm}, {Wittenmyer},
  {Olivares E.}, {Melo}, {Rojo}, {Jord{\'a}n}, {Drass}, {Butler}, \&
  {Wang}}]{Jones_2016}
{Jones}, M.~I., {Jenkins}, J.~S., {Brahm}, R., {et~al.} 2016,
  \href{http://dx.doi.org/10.1051/0004-6361/201628067}{\color{magenta}\aap},
  \href{https://ui.adsabs.harvard.edu/abs/2016A&A...590A..38J}{\color{cyan}590},
  A38

\bibitem[{{Kennedy} \& {Kenyon}(2008)}]{Kennedy_2008}
{Kennedy}, G.~M., \& {Kenyon}, S.~J. 2008,
  \href{http://dx.doi.org/10.1086/524130}{\color{magenta}\apj},
  \href{https://ui.adsabs.harvard.edu/abs/2008ApJ...673..502K}{\color{cyan}673},
  502

\bibitem[{{Kennedy} \& {Kenyon}(2009)}]{Kennedy_2009}
---. 2009,
  \href{http://dx.doi.org/10.1088/0004-637X/695/2/1210}{\color{magenta}\apj},
  \href{https://ui.adsabs.harvard.edu/abs/2009ApJ...695.1210K}{\color{cyan}695},
  1210

\bibitem[{{Kilic} {et~al.}(2021){Kilic}, {Bergeron}, {Blouin}, \&
  {B{\'e}dard}}]{Kilic_2021_massive}
{Kilic}, M., {Bergeron}, P., {Blouin}, S., \& {B{\'e}dard}, A. 2021,
  \href{http://dx.doi.org/10.1093/mnras/stab767}{\color{magenta}\mnras},
  \href{https://ui.adsabs.harvard.edu/abs/2021MNRAS.503.5397K}{\color{cyan}503},
  5397

\bibitem[{{Kilic} {et~al.}(2009){Kilic}, {Gould}, \& {Koester}}]{Kilic_2009}
{Kilic}, M., {Gould}, A., \& {Koester}, D. 2009,
  \href{http://dx.doi.org/10.1088/0004-637X/705/2/1219}{\color{magenta}\apj},
  \href{https://ui.adsabs.harvard.edu/abs/2009ApJ...705.1219K}{\color{cyan}705},
  1219

\bibitem[{{Kilic} {et~al.}(2023){Kilic}, {Moss}, {Kosakowski}, {Bergeron},
  {Conly}, {Brown}, {Toonen}, {Williams}, \& {Dufour}}]{Kilic_2023}
{Kilic}, M., {Moss}, A.~G., {Kosakowski}, A., {et~al.} 2023,
  \href{http://dx.doi.org/10.1093/mnras/stac3182}{\color{magenta}\mnras},
  \href{https://ui.adsabs.harvard.edu/abs/2023MNRAS.518.2341K}{\color{cyan}518},
  2341

\bibitem[{{Kowalski} \& {Saumon}(2006)}]{Kowalski_2006}
{Kowalski}, P.~M., \& {Saumon}, D. 2006,
  \href{http://dx.doi.org/10.1086/509723}{\color{magenta}\apjl},
  \href{https://ui.adsabs.harvard.edu/abs/2006ApJ...651L.137K}{\color{cyan}651},
  L137

\bibitem[{{Kretke} {et~al.}(2009){Kretke}, {Lin}, {Garaud}, \&
  {Turner}}]{Kretke_2009}
{Kretke}, K.~A., {Lin}, D.~N.~C., {Garaud}, P., \& {Turner}, N.~J. 2009,
  \href{http://dx.doi.org/10.1088/0004-637X/690/1/407}{\color{magenta}\apj},
  \href{https://ui.adsabs.harvard.edu/abs/2009ApJ...690..407K}{\color{cyan}690},
  407

\bibitem[{{Lai} {et~al.}(2021){Lai}, {Dennihy}, {Xu}, {Nitta}, {Kleinman},
  {Leggett}, {Bonsor}, {Hodgkin}, {Rebassa-Mansergas}, \& {Rogers}}]{Lai_2021}
{Lai}, S., {Dennihy}, E., {Xu}, S., {et~al.} 2021,
  \href{http://dx.doi.org/10.3847/1538-4357/ac1354}{\color{magenta}\apj},
  \href{https://ui.adsabs.harvard.edu/abs/2021ApJ...920..156L}{\color{cyan}920},
  156

\bibitem[{{Laughlin} {et~al.}(2004){Laughlin}, {Bodenheimer}, \&
  {Adams}}]{Laughlin_2004}
{Laughlin}, G., {Bodenheimer}, P., \& {Adams}, F.~C. 2004,
  \href{http://dx.doi.org/10.1086/424384}{\color{magenta}\apjl},
  \href{https://ui.adsabs.harvard.edu/abs/2004ApJ...612L..73L}{\color{cyan}612},
  L73

\bibitem[{{Limbach} {et~al.}(2024){Limbach}, {Vanderburg}, {Venner}, {Blouin},
  {Stevenson}, {MacDonald}, {Jenkins}, {Bowens-Rubin}, {Soares-Furtado},
  {Morley}, {Janson}, {Debes}, {Xu}, {Kleisioti}, {Kenworthy}, {Butler},
  {Crane}, {Osip}, {Shectman}, \& {Teske}}]{Limbach_2024}
{Limbach}, M.~A., {Vanderburg}, A., {Venner}, A., {et~al.} 2024,
  \href{http://dx.doi.org/10.3847/2041-8213/ad74ed}{\color{magenta}\apjl},
  \href{https://ui.adsabs.harvard.edu/abs/2024ApJ...973L..11L}{\color{cyan}973},
  L11

\bibitem[{{Lindegren} {et~al.}(2021{\natexlab{a}}){Lindegren}, {Bastian},
  {Biermann}, {Bombrun}, {de Torres}, {Gerlach}, {Geyer}, {Hern{\'a}ndez},
  {Hilger}, {Hobbs}, {Klioner}, {Lammers}, {McMillan}, {Ramos-Lerate},
  {Steidelm{\"u}ller}, {Stephenson}, \& {van Leeuwen}}]{Lindegren_2021a}
{Lindegren}, L., {Bastian}, U., {Biermann}, M., {et~al.} 2021{\natexlab{a}},
  \href{http://dx.doi.org/10.1051/0004-6361/202039653}{\color{magenta}\aap},
  \href{https://ui.adsabs.harvard.edu/abs/2021A&A...649A...4L}{\color{cyan}649},
  A4

\bibitem[{{Lindegren} {et~al.}(2021{\natexlab{b}}){Lindegren}, {Klioner},
  {Hern{\'a}ndez}, {Bombrun}, {Ramos-Lerate}, {Steidelm{\"u}ller}, {Bastian},
  {Biermann}, {de Torres}, {Gerlach}, {Geyer}, {Hilger}, {Hobbs}, {Lammers},
  {McMillan}, {Stephenson}, {Casta{\~n}eda}, {Davidson}, {Fabricius},
  {Gracia-Abril}, {Portell}, {Rowell}, {Teyssier}, {Torra}, {Bartolom{\'e}},
  {Clotet}, {Garralda}, {Gonz{\'a}lez-Vidal}, {Torra}, {Abbas}, {Altmann},
  {Anglada Varela}, {Balaguer-N{\'u}{\~n}ez}, {Balog}, {Barache}, {Becciani},
  {Bernet}, {Bertone}, {Bianchi}, {Bouquillon}, {Brown}, {Bucciarelli},
  {Busonero}, {Butkevich}, {Buzzi}, {Cancelliere}, {Carlucci}, {Charlot},
  {Cioni}, {Crosta}, {Crowley}, {del Peloso}, {del Pozo}, {Drimmel}, {Esquej},
  {Fienga}, {Fraile}, {Gai}, {Garcia-Reinaldos}, {Guerra}, {Hambly}, {Hauser},
  {Jan{\ss}en}, {Jordan}, {Kostrzewa-Rutkowska}, {Lattanzi}, {Liao}, {Licata},
  {Lister}, {L{\"o}ffler}, {Marchant}, {Masip}, {Mignard}, {Mints}, {Molina},
  {Mora}, {Morbidelli}, {Murphy}, {Pagani}, {Panuzzo}, {Pe{\~n}alosa Esteller},
  {Poggio}, {Re Fiorentin}, {Riva}, {Sagrist{\`a} Sell{\'e}s}, {Sanchez
  Gimenez}, {Sarasso}, {Sciacca}, {Siddiqui}, {Smart}, {Souami}, {Spagna},
  {Steele}, {Taris}, {Utrilla}, {van Reeven}, \& {Vecchiato}}]{Lindegren_2021b}
{Lindegren}, L., {Klioner}, S.~A., {Hern{\'a}ndez}, J., {et~al.}
  2021{\natexlab{b}},
  \href{http://dx.doi.org/10.1051/0004-6361/202039709}{\color{magenta}\aap},
  \href{https://ui.adsabs.harvard.edu/abs/2021A&A...649A...2L}{\color{cyan}649},
  A2

\bibitem[{{Mainzer} {et~al.}(2011){Mainzer}, {Grav}, {Bauer}, {Masiero},
  {McMillan}, {Cutri}, {Walker}, {Wright}, {Eisenhardt}, {Tholen}, {Spahr},
  {Jedicke}, {Denneau}, {DeBaun}, {Elsbury}, {Gautier}, {Gomillion}, {Hand},
  {Mo}, {Watkins}, {Wilkins}, {Bryngelson}, {Del Pino Molina}, {Desai},
  {G{\'o}mez Camus}, {Hidalgo}, {Konstantopoulos}, {Larsen}, {Maleszewski},
  {Malkan}, {Mauduit}, {Mullan}, {Olszewski}, {Pforr}, {Saro}, {Scotti}, \&
  {Wasserman}}]{Mainzer_2011}
{Mainzer}, A., {Grav}, T., {Bauer}, J., {et~al.} 2011,
  \href{http://dx.doi.org/10.1088/0004-637X/743/2/156}{\color{magenta}\apj},
  \href{https://ui.adsabs.harvard.edu/abs/2011ApJ...743..156M}{\color{cyan}743},
  156

\bibitem[{{Maldonado} {et~al.}(2013){Maldonado}, {Villaver}, \&
  {Eiroa}}]{Maldonado_2013}
{Maldonado}, J., {Villaver}, E., \& {Eiroa}, C. 2013,
  \href{http://dx.doi.org/10.1051/0004-6361/201321082}{\color{magenta}\aap},
  \href{https://ui.adsabs.harvard.edu/abs/2013A&A...554A..84M}{\color{cyan}554},
  A84

\bibitem[{{Marley} {et~al.}(2021){Marley}, {Saumon}, {Visscher}, {Lupu},
  {Freedman}, {Morley}, {Fortney}, {Seay}, {Smith}, {Teal}, \&
  {Wang}}]{Marley_2021}
{Marley}, M.~S., {Saumon}, D., {Visscher}, C., {et~al.} 2021,
  \href{http://dx.doi.org/10.3847/1538-4357/ac141d}{\color{magenta}\apj},
  \href{https://ui.adsabs.harvard.edu/abs/2021ApJ...920...85M}{\color{cyan}920},
  85

\bibitem[{{Marocco} {et~al.}(2021){Marocco}, {Eisenhardt}, {Fowler},
  {Kirkpatrick}, {Meisner}, {Schlafly}, {Stanford}, {Garcia}, {Caselden},
  {Cushing}, {Cutri}, {Faherty}, {Gelino}, {Gonzalez}, {Jarrett}, {Koontz},
  {Mainzer}, {Marchese}, {Mobasher}, {Schlegel}, {Stern}, {Teplitz}, \&
  {Wright}}]{Marocco_2021}
{Marocco}, F., {Eisenhardt}, P. R.~M., {Fowler}, J.~W., {et~al.} 2021,
  \href{http://dx.doi.org/10.3847/1538-4365/abd805}{\color{magenta}\apjs},
  \href{https://ui.adsabs.harvard.edu/abs/2021ApJS..253....8M}{\color{cyan}253},
  8

\bibitem[{{McMahon} {et~al.}(2013){McMahon}, {Banerji}, {Gonzalez}, {Koposov},
  {Bejar}, {Lodieu}, {Rebolo}, \& {VHS Collaboration}}]{McMahon_2013}
{McMahon}, R.~G., {Banerji}, M., {Gonzalez}, E., {et~al.} 2013, The Messenger,
  \href{https://ui.adsabs.harvard.edu/abs/2013Msngr.154...35M}{\color{cyan}154},
  35

\bibitem[{{Mullally} {et~al.}(2007){Mullally}, {Kilic}, {Reach}, {Kuchner},
  {von Hippel}, {Burrows}, \& {Winget}}]{Mullally_2007}
{Mullally}, F., {Kilic}, M., {Reach}, W.~T., {et~al.} 2007,
  \href{http://dx.doi.org/10.1086/511858}{\color{magenta}\apjs},
  \href{https://ui.adsabs.harvard.edu/abs/2007ApJS..171..206M}{\color{cyan}171},
  206

\bibitem[{{Mullally} {et~al.}(2024){Mullally}, {Debes}, {Cracraft}, {Mullally},
  {Poulsen}, {Albert}, {Thibault}, {Reach}, {Hermes}, {Barclay}, {Kilic}, \&
  {Quintana}}]{Mullally_2024}
{Mullally}, S.~E., {Debes}, J., {Cracraft}, M., {et~al.} 2024,
  \href{http://dx.doi.org/10.3847/2041-8213/ad2348}{\color{magenta}\apjl},
  \href{https://ui.adsabs.harvard.edu/abs/2024ApJ...962L..32M}{\color{cyan}962},
  L32

\bibitem[{{Mustill} \& {Villaver}(2012)}]{Mustill_2012}
{Mustill}, A.~J., \& {Villaver}, E. 2012,
  \href{http://dx.doi.org/10.1088/0004-637X/761/2/121}{\color{magenta}\apj},
  \href{https://ui.adsabs.harvard.edu/abs/2012ApJ...761..121M}{\color{cyan}761},
  121

\bibitem[{{Nielsen} {et~al.}(2019){Nielsen}, {De Rosa}, {Macintosh}, {Wang},
  {Ruffio}, {Chiang}, {Marley}, {Saumon}, {Savransky}, {Ammons}, {Bailey},
  {Barman}, {Blain}, {Bulger}, {Burrows}, {Chilcote}, {Cotten}, {Czekala},
  {Doyon}, {Duch{\^e}ne}, {Esposito}, {Fabrycky}, {Fitzgerald}, {Follette},
  {Fortney}, {Gerard}, {Goodsell}, {Graham}, {Greenbaum}, {Hibon}, {Hinkley},
  {Hirsch}, {Hom}, {Hung}, {Dawson}, {Ingraham}, {Kalas}, {Konopacky},
  {Larkin}, {Lee}, {Lin}, {Maire}, {Marchis}, {Marois}, {Metchev},
  {Millar-Blanchaer}, {Morzinski}, {Oppenheimer}, {Palmer}, {Patience},
  {Perrin}, {Poyneer}, {Pueyo}, {Rafikov}, {Rajan}, {Rameau}, {Rantakyr{\"o}},
  {Ren}, {Schneider}, {Sivaramakrishnan}, {Song}, {Soummer}, {Tallis},
  {Thomas}, {Ward-Duong}, \& {Wolff}}]{Nielsen_2019}
{Nielsen}, E.~L., {De Rosa}, R.~J., {Macintosh}, B., {et~al.} 2019,
  \href{http://dx.doi.org/10.3847/1538-3881/ab16e9}{\color{magenta}\aj},
  \href{https://ui.adsabs.harvard.edu/abs/2019AJ....158...13N}{\color{cyan}158},
  13

\bibitem[{{Pasquini} {et~al.}(2007){Pasquini}, {D{\"o}llinger}, {Weiss},
  {Girardi}, {Chavero}, {Hatzes}, {da Silva}, \& {Setiawan}}]{Pasquini_2007}
{Pasquini}, L., {D{\"o}llinger}, M.~P., {Weiss}, A., {et~al.} 2007,
  \href{http://dx.doi.org/10.1051/0004-6361:20077814}{\color{magenta}\aap},
  \href{https://ui.adsabs.harvard.edu/abs/2007A&A...473..979P}{\color{cyan}473},
  979

\bibitem[{{Phillips} {et~al.}(2020){Phillips}, {Tremblin}, {Baraffe},
  {Chabrier}, {Allard}, {Spiegelman}, {Goyal}, {Drummond}, \&
  {H{\'e}brard}}]{Phillips_2020}
{Phillips}, M.~W., {Tremblin}, P., {Baraffe}, I., {et~al.} 2020,
  \href{http://dx.doi.org/10.1051/0004-6361/201937381}{\color{magenta}\aap},
  \href{https://ui.adsabs.harvard.edu/abs/2020A&A...637A..38P}{\color{cyan}637},
  A38

\bibitem[{{Pollack} {et~al.}(1996){Pollack}, {Hubickyj}, {Bodenheimer},
  {Lissauer}, {Podolak}, \& {Greenzweig}}]{Pollack_1996}
{Pollack}, J.~B., {Hubickyj}, O., {Bodenheimer}, P., {et~al.} 1996,
  \href{http://dx.doi.org/10.1006/icar.1996.0190}{\color{magenta}\icarus},
  \href{https://ui.adsabs.harvard.edu/abs/1996Icar..124...62P}{\color{cyan}124},
  62

\bibitem[{{R Core Team}(2024)}]{R}
{R Core Team}. 2024, R: A Language and Environment for Statistical Computing, R
  Foundation for Statistical Computing, Vienna, Austria

\bibitem[{{Reffert} {et~al.}(2015){Reffert}, {Bergmann}, {Quirrenbach},
  {Trifonov}, \& {K{\"u}nstler}}]{Reffert_2015}
{Reffert}, S., {Bergmann}, C., {Quirrenbach}, A., {Trifonov}, T., \&
  {K{\"u}nstler}, A. 2015,
  \href{http://dx.doi.org/10.1051/0004-6361/201322360}{\color{magenta}\aap},
  \href{https://ui.adsabs.harvard.edu/abs/2015A&A...574A.116R}{\color{cyan}574},
  A116

\bibitem[{{Ribas} {et~al.}(2015){Ribas}, {Bouy}, \& {Mer{\'\i}n}}]{Ribas_2015}
{Ribas}, {\'A}., {Bouy}, H., \& {Mer{\'\i}n}, B. 2015,
  \href{http://dx.doi.org/10.1051/0004-6361/201424846}{\color{magenta}\aap},
  \href{https://ui.adsabs.harvard.edu/abs/2015A&A...576A..52R}{\color{cyan}576},
  A52

\bibitem[{{Riello} {et~al.}(2021){Riello}, {De Angeli}, {Evans}, {Montegriffo},
  {Carrasco}, {Busso}, {Palaversa}, {Burgess}, {Diener}, {Davidson}, {Rowell},
  {Fabricius}, {Jordi}, {Bellazzini}, {Pancino}, {Harrison}, {Cacciari}, {van
  Leeuwen}, {Hambly}, {Hodgkin}, {Osborne}, {Altavilla}, {Barstow}, {Brown},
  {Castellani}, {Cowell}, {De Luise}, {Gilmore}, {Giuffrida}, {Hidalgo},
  {Holland}, {Marinoni}, {Pagani}, {Piersimoni}, {Pulone}, {Ragaini}, {Rainer},
  {Richards}, {Sanna}, {Walton}, {Weiler}, \& {Yoldas}}]{Riello_2021}
{Riello}, M., {De Angeli}, F., {Evans}, D.~W., {et~al.} 2021,
  \href{http://dx.doi.org/10.1051/0004-6361/202039587}{\color{magenta}\aap},
  \href{https://ui.adsabs.harvard.edu/abs/2021A&A...649A...3R}{\color{cyan}649},
  A3

\bibitem[{{Roccatagliata} {et~al.}(2011){Roccatagliata}, {Bouwman}, {Henning},
  {Gennaro}, {Feigelson}, {Kim}, {Sicilia-Aguilar}, \&
  {Lawson}}]{Roccatagliata_2011}
{Roccatagliata}, V., {Bouwman}, J., {Henning}, T., {et~al.} 2011,
  \href{http://dx.doi.org/10.1088/0004-637X/733/2/113}{\color{magenta}\apj},
  \href{https://ui.adsabs.harvard.edu/abs/2011ApJ...733..113R}{\color{cyan}733},
  113

\bibitem[{{Rowell} {et~al.}(2021){Rowell}, {Davidson}, {Lindegren}, {van
  Leeuwen}, {Casta{\~n}eda}, {Fabricius}, {Bastian}, {Hambly}, {Hern{\'a}ndez},
  {Bombrun}, {Evans}, {De Angeli}, {Riello}, {Busonero}, {Crowley}, {Mora},
  {Lammers}, {Gracia}, {Portell}, {Biermann}, \& {Brown}}]{Rowell_2021}
{Rowell}, N., {Davidson}, M., {Lindegren}, L., {et~al.} 2021,
  \href{http://dx.doi.org/10.1051/0004-6361/202039448}{\color{magenta}\aap},
  \href{https://ui.adsabs.harvard.edu/abs/2021A&A...649A..11R}{\color{cyan}649},
  A11

\bibitem[{{Salaris} {et~al.}(2013){Salaris}, {Althaus}, \&
  {Garc{\'\i}a-Berro}}]{Salaris_2013}
{Salaris}, M., {Althaus}, L.~G., \& {Garc{\'\i}a-Berro}, E. 2013,
  \href{http://dx.doi.org/10.1051/0004-6361/201220622}{\color{magenta}\aap},
  \href{https://ui.adsabs.harvard.edu/abs/2013A&A...555A..96S}{\color{cyan}555},
  A96

\bibitem[{{Sanders} {et~al.}(2007){Sanders}, {Salvato}, {Aussel}, {Ilbert},
  {Scoville}, {Surace}, {Frayer}, {Sheth}, {Helou}, {Brooke}, {Bhattacharya},
  {Yan}, {Kartaltepe}, {Barnes}, {Blain}, {Calzetti}, {Capak}, {Carilli},
  {Carollo}, {Comastri}, {Daddi}, {Ellis}, {Elvis}, {Fall}, {Franceschini},
  {Giavalisco}, {Hasinger}, {Impey}, {Koekemoer}, {Le F{\`e}vre}, {Lilly},
  {Liu}, {McCracken}, {Mobasher}, {Renzini}, {Rich}, {Schinnerer}, {Shopbell},
  {Taniguchi}, {Thompson}, {Urry}, \& {Williams}}]{Sanders_2007}
{Sanders}, D.~B., {Salvato}, M., {Aussel}, H., {et~al.} 2007,
  \href{http://dx.doi.org/10.1086/517885}{\color{magenta}\apjs},
  \href{https://ui.adsabs.harvard.edu/abs/2007ApJS..172...86S}{\color{cyan}172},
  86

\bibitem[{{Santos} {et~al.}(2004){Santos}, {Israelian}, \&
  {Mayor}}]{Santos_2004}
{Santos}, N.~C., {Israelian}, G., \& {Mayor}, M. 2004,
  \href{http://dx.doi.org/10.1051/0004-6361:20034469}{\color{magenta}\aap},
  \href{https://ui.adsabs.harvard.edu/abs/2004A&A...415.1153S}{\color{cyan}415},
  1153

\bibitem[{{Schlaufman}(2014)}]{Schlaufman_2014}
{Schlaufman}, K.~C. 2014,
  \href{http://dx.doi.org/10.1088/0004-637X/790/2/91}{\color{magenta}\apj},
  \href{https://ui.adsabs.harvard.edu/abs/2014ApJ...790...91S}{\color{cyan}790},
  91

\bibitem[{{Squicciarini} {et~al.}(2022){Squicciarini}, {Gratton}, {Janson},
  {Mamajek}, {Chauvin}, {Delorme}, {Langlois}, {Vigan}, {Ringqvist}, {Meeus},
  {Reffert}, {Kenworthy}, {Meyer}, {Bonnefoy}, {Bonavita}, {Mesa}, {Samland},
  {Desidera}, {D'Orazi}, {Engler}, {Alecian}, {Miglio}, {Henning}, {Quanz},
  {Mayer}, {Flasseur}, \& {Marleau}}]{Squicciarini_2022}
{Squicciarini}, V., {Gratton}, R., {Janson}, M., {et~al.} 2022,
  \href{http://dx.doi.org/10.1051/0004-6361/202243675}{\color{magenta}\aap},
  \href{https://ui.adsabs.harvard.edu/abs/2022A&A...664A...9S}{\color{cyan}664},
  A9

\bibitem[{{SSC And IRSA}(2020)}]{https://doi.org/10.26131/irsa433}
{SSC And IRSA}. 2020, Spitzer Enhanced Imaging Products,  IPAC

\bibitem[{{Takeda} {et~al.}(2008){Takeda}, {Sato}, \& {Murata}}]{Takeda_2008}
{Takeda}, Y., {Sato}, B., \& {Murata}, D. 2008,
  \href{http://dx.doi.org/10.1093/pasj/60.4.781}{\color{magenta}\pasj},
  \href{https://ui.adsabs.harvard.edu/abs/2008PASJ...60..781T}{\color{cyan}60},
  781

\bibitem[{{Torra} {et~al.}(2021){Torra}, {Casta{\~n}eda}, {Fabricius},
  {Lindegren}, {Clotet}, {Gonz{\'a}lez-Vidal}, {Bartolom{\'e}}, {Bastian},
  {Bernet}, {Biermann}, {Garralda}, {G{\'u}rpide}, {Lammers}, {Portell}, \&
  {Torra}}]{Torra_2021}
{Torra}, F., {Casta{\~n}eda}, J., {Fabricius}, C., {et~al.} 2021,
  \href{http://dx.doi.org/10.1051/0004-6361/202039637}{\color{magenta}\aap},
  \href{https://ui.adsabs.harvard.edu/abs/2021A&A...649A..10T}{\color{cyan}649},
  A10

\bibitem[{{Tremblay} {et~al.}(2011){Tremblay}, {Bergeron}, \&
  {Gianninas}}]{Tremblay_2011}
{Tremblay}, P.~E., {Bergeron}, P., \& {Gianninas}, A. 2011,
  \href{http://dx.doi.org/10.1088/0004-637X/730/2/128}{\color{magenta}\apj},
  \href{https://ui.adsabs.harvard.edu/abs/2011ApJ...730..128T}{\color{cyan}730},
  128

\bibitem[{{Veras} {et~al.}(2020){Veras}, {Tremblay}, {Hermes}, {McDonald},
  {Kennedy}, {Meru}, \& {G{\"a}nsicke}}]{Veras_2020}
{Veras}, D., {Tremblay}, P.-E., {Hermes}, J.~J., {et~al.} 2020,
  \href{http://dx.doi.org/10.1093/mnras/staa241}{\color{magenta}\mnras},
  \href{https://ui.adsabs.harvard.edu/abs/2020MNRAS.493..765V}{\color{cyan}493},
  765

\bibitem[{{Veras} {et~al.}(2011){Veras}, {Wyatt}, {Mustill}, {Bonsor}, \&
  {Eldridge}}]{Veras_2011}
{Veras}, D., {Wyatt}, M.~C., {Mustill}, A.~J., {Bonsor}, A., \& {Eldridge},
  J.~J. 2011,
  \href{http://dx.doi.org/10.1111/j.1365-2966.2011.19393.x}{\color{magenta}\mnras},
  \href{https://ui.adsabs.harvard.edu/abs/2011MNRAS.417.2104V}{\color{cyan}417},
  2104

\bibitem[{Virtanen {et~al.}(2020)Virtanen, Gommers, Oliphant, Haberland, Reddy,
  Cournapeau, Burovski, Peterson, Weckesser, Bright, {van der Walt}, Brett,
  Wilson, Millman, Mayorov, Nelson, Jones, Kern, Larson, Carey, Polat, Feng,
  Moore, {VanderPlas}, Laxalde, Perktold, Cimrman, Henriksen, Quintero, Harris,
  Archibald, Ribeiro, Pedregosa, {van Mulbregt}, \& {SciPy 1.0
  Contributors}}]{scipy}
Virtanen, P., Gommers, R., Oliphant, T.~E., {et~al.} 2020,
  \href{http://dx.doi.org/10.1038/s41592-019-0686-2}{\color{magenta}Nature
  Methods}, \href{https://rdcu.be/b08Wh}{\color{cyan}17}, 261

\bibitem[{{Wilson} {et~al.}(2019){Wilson}, {Farihi}, {G{\"a}nsicke}, \&
  {Swan}}]{Wilson_2019}
{Wilson}, T.~G., {Farihi}, J., {G{\"a}nsicke}, B.~T., \& {Swan}, A. 2019,
  \href{http://dx.doi.org/10.1093/mnras/stz1050}{\color{magenta}\mnras},
  \href{https://ui.adsabs.harvard.edu/abs/2019MNRAS.487..133W}{\color{cyan}487},
  133

\bibitem[{{Wittenmyer} {et~al.}(2017){Wittenmyer}, {Jones}, {Zhao}, {Marshall},
  {Butler}, {Tinney}, {Wang}, \& {Johnson}}]{Wittenmyer_2017}
{Wittenmyer}, R.~A., {Jones}, M.~I., {Zhao}, J., {et~al.} 2017,
  \href{http://dx.doi.org/10.3847/1538-3881/153/2/51}{\color{magenta}\aj},
  \href{https://ui.adsabs.harvard.edu/abs/2017AJ....153...51W}{\color{cyan}153},
  51

\bibitem[{{Wolthoff} {et~al.}(2022){Wolthoff}, {Reffert}, {Quirrenbach},
  {Jones}, {Wittenmyer}, \& {Jenkins}}]{Wolthoff_2022}
{Wolthoff}, V., {Reffert}, S., {Quirrenbach}, A., {et~al.} 2022,
  \href{http://dx.doi.org/10.1051/0004-6361/202142501}{\color{magenta}\aap},
  \href{https://ui.adsabs.harvard.edu/abs/2022A&A...661A..63W}{\color{cyan}661},
  A63

\bibitem[{{Wright} {et~al.}(2010){Wright}, {Eisenhardt}, {Mainzer}, {Ressler},
  {Cutri}, {Jarrett}, {Kirkpatrick}, {Padgett}, {McMillan}, {Skrutskie},
  {Stanford}, {Cohen}, {Walker}, {Mather}, {Leisawitz}, {Gautier}, {McLean},
  {Benford}, {Lonsdale}, {Blain}, {Mendez}, {Irace}, {Duval}, {Liu}, {Royer},
  {Heinrichsen}, {Howard}, {Shannon}, {Kendall}, {Walsh}, {Larsen}, {Cardon},
  {Schick}, {Schwalm}, {Abid}, {Fabinsky}, {Naes}, \& {Tsai}}]{Wright_2010}
{Wright}, E.~L., {Eisenhardt}, P. R.~M., {Mainzer}, A.~K., {et~al.} 2010,
  \href{http://dx.doi.org/10.1088/0004-6256/140/6/1868}{\color{magenta}\aj},
  \href{https://ui.adsabs.harvard.edu/abs/2010AJ....140.1868W}{\color{cyan}140},
  1868

\bibitem[{{Xu} {et~al.}(2015){Xu}, {Ertel}, {Wahhaj}, {Milli}, {Scicluna}, \&
  {Bertrang}}]{Xu_2015}
{Xu}, S., {Ertel}, S., {Wahhaj}, Z., {et~al.} 2015,
  \href{http://dx.doi.org/10.1051/0004-6361/201526179}{\color{magenta}\aap},
  \href{https://ui.adsabs.harvard.edu/abs/2015A&A...579L...8X}{\color{cyan}579},
  L8

\bibitem[{{Xu} {et~al.}(2020){Xu}, {Lai}, \& {Dennihy}}]{Xu_2020}
{Xu}, S., {Lai}, S., \& {Dennihy}, E. 2020,
  \href{http://dx.doi.org/10.3847/1538-4357/abb3fc}{\color{magenta}\apj},
  \href{https://ui.adsabs.harvard.edu/abs/2020ApJ...902..127X}{\color{cyan}902},
  127

\bibitem[{{Yasui} {et~al.}(2014){Yasui}, {Kobayashi}, {Tokunaga}, \&
  {Saito}}]{Yasui_2014}
{Yasui}, C., {Kobayashi}, N., {Tokunaga}, A.~T., \& {Saito}, M. 2014,
  \href{http://dx.doi.org/10.1093/mnras/stu1013}{\color{magenta}\mnras},
  \href{https://ui.adsabs.harvard.edu/abs/2014MNRAS.442.2543Y}{\color{cyan}442},
  2543

\end{thebibliography}
\bibliographystyle{aasjournal_sihao}
\end{CJK*}
\end{document}